\documentclass[11pt]{article}

\usepackage[utf8]{inputenc}
\usepackage[T1]{fontenc}
\usepackage{lmodern}
\usepackage{amsmath,amssymb,amsthm}
\usepackage{booktabs}
\usepackage{array}
\usepackage{graphicx}
\usepackage{hyperref}
\usepackage{achicago}
\usepackage{url}
\usepackage{enumitem}
\usepackage{longtable}
\usepackage{tabularx}
\usepackage{caption}
\usepackage{float}
\usepackage{placeins}
\usepackage{pgfplots}
\usepackage{pgfplotstable}
\pgfplotsset{compat=1.18}
\providecommand{\citep}[1]{\cite{#1}}
\providecommand{\shorttitle}[1]{}
\providecommand{\runningtitle}[1]{}
\newcolumntype{P}[1]{>{\raggedright\arraybackslash}p{#1}}
\newcolumntype{Y}{>{\centering\arraybackslash}X}

\hypersetup{
  colorlinks=true,
  linkcolor=blue,
  citecolor=blue,
  urlcolor=blue,
  pdftitle={Regenerative Bonds: Formal Debt, Mutual-Aid, and Local Settlement Capacity},
  pdfauthor={William O. Ruddick; Alex (Yeshey) Cahana; Tom Shael}
}

\newtheorem{definition}{Definition}

\makeatletter
\renewcommand{\maketitle}{%
  \begingroup
  \begin{center}
    \vspace*{-2.6em}
    {\Large\bfseries \@title\par}
    \vspace{0.6em}
    {\normalsize
    William O. Ruddick\\
    Grassroots Economics Foundation, Kilifi, Kenya\\[0.25em]
    Alex (Yeshey) Cahana\\
    Drinchengang Regenerative Village, Bhutan\\[0.25em]
    Tom Shael\\
    Independent Researcher\par}
    \vspace{0.35em}
    {\normalsize \@date\par}
    \vspace{0.35em}
    {\footnotesize Corresponding author: \texttt{will@grassecon.org}\par}
  \end{center}
  \vspace{-0.9em}
  \endgroup
}
\renewenvironment{abstract}{%
  \begin{center}
  \bfseries Abstract
  \end{center}
  \vspace{-0.6em}
  \small
  \noindent
}{\par\vspace{0.2em}}
\makeatother

\title{Regenerative Bonds: Formal Debt, Mutual-Aid, and Local Settlement Capacity}
\shorttitle{Regenerative Bonds}
\runningtitle{Regenerative Bonds}
\author{William O. Ruddick \\
Grassroots Economics Foundation, Kilifi, Kenya \\
\and
Alex (Yeshey) Cahana \\
Drinchengang Regenerative Village, Bhutan \\
\and
Tom Shael\\
Independent Researcher}
\date{June 2026}

\begin{document}
\maketitle

\begin{abstract}
This paper develops regenerative bonds as formal debt instruments whose disclosed use-of-proceeds and governance rules allocate proceeds to locally governed settlement systems designed to strengthen settlement capacity across locally specified productive, ecological, care, mutual-aid, and repair commitments without converting those commitments into investor collateral. It separates bondholder claims from local redeemable commitments and models commitment pools that curate, value, limit, exchange, route, and repair those commitments. Sarafu Network, based in Kenya, provides component evidence on commitment circulation, stable-value interaction, liquidity, topology, and report-linked activity. A Monte Carlo engine calibrated to privacy-safe empirical moments asks whether bond liquidity can act as reusable catalytic funding while preserving issuer responsibility for debt service. Under the reported assumptions, the frontier identifies a modeled guardrail-pass region in which scheduled service is preserved, mutual-aid circulation is maintained or amplified, and bond issuer headroom remains available in lower-stress cells; edge diagnostics show that higher debt-service pressure and capital intensity narrow this region. The contribution is a settlement-architecture framework for evaluating when formal debt can strengthen local capacity to fulfill and repair commitments without becoming hidden household collateral.
\end{abstract}


\newpage

\section{Introduction}
Bonds can mobilize capital across time, but conventional bond architecture says relatively little about how financed activity circulates after proceeds are disbursed. A green or social bond can specify eligible uses of proceeds and reporting while local procurement, mutual-aid institutions, settlement bottlenecks, and repair paths remain unchanged. Here settlement architecture means the institutional design of who owes what, through which instruments, under which governance rules, and with what repair paths. This paper asks whether formal debt can support locally governed capacity to fulfill and repair commitments without converting those commitments into investor collateral.

The institutional boundary is central. A formal bond is a claim on an issuer that may be transferable under instrument terms and applicable law. A local redeemable commitment is a local promise to provide specified fulfillment, redemption, or repair under governed terms; it may circulate only where terms and law allow. Both are obligation instruments, but they do not create the same legal claim, creditor class, or enforcement path.

This paper defines a regenerative bond as a formal debt instrument whose disclosed use-of-proceeds and governance rules allocate proceeds to locally governed settlement systems designed to strengthen settlement capacity across locally specified productive, ecological, care, mutual-aid, and repair commitments while preserving issuer responsibility for scheduled obligations to bondholders. The testable mechanism is conditional: federated commitment pools may increase route availability and obligation-closing settlement enough to create eligible service capacity under stated governance and non-extraction constraints. The empirical and simulation sections test settlement-capacity conditions rather than investment viability; causal welfare, cultural-continuity, and ecological claims remain outside the identified evidence.

\paragraph{Contribution}
The paper makes four contributions. First, it defines regenerative bonds as formal debt instruments that allocate proceeds to locally governed settlement systems while preserving issuer responsibility for scheduled obligations to bondholders. Second, it specifies the non-extraction boundary between formal bondholder claims and local redeemable commitments. Third, it uses Commitment Pooling Protocol (CPP) as an operational grammar for curation, valuation, limitation, exchange, routing, settlement, and repair. Fourth, it combines Sarafu Network component evidence with Monte Carlo stress testing to evaluate whether repayable liquidity can support local settlement without becoming hidden household collateral. Cultural-continuity language is bounded to analytical motifs grounded in the cited literature and practitioner record.

\paragraph{Evidence status}
Sarafu Network provides component-level implementation and calibration evidence. It shows that several mechanisms required by the framework have been implemented and measured in practice, while leaving the formal bond layer as a counterfactual simulation \citep{ruddick2026sarafuDataset}. Transaction data, verified reports, and audit summaries provide evidence of reported activities, pool interaction, fulfillment records, and calibration moments; they do not by themselves establish causal welfare effects, legal compliance, consent, cultural legitimacy, or universal ecological regeneration. Those stronger claims require verification-tier coding, legal review, and, where appropriate, causal or quasi-experimental designs.

This paper evaluates mechanism plausibility and frontier conditions, not investment suitability or policy readiness.

\section{Bond Finance and the Settlement-Architecture Gap}
Formal bonds are central instruments of public finance, development finance, climate-transition investment, and corporate capital formation. Public-debt management literature emphasizes that bond issuance creates direct and contingent obligations that must remain visible to fiscal authorities and investors \citep{oecd2025globaldebt,worldbankimf2014debt}. A regenerative bond does not remove the issuer's debt-service obligation, refinancing risk, interest-rate exposure, or disclosure duties. Its regenerative claim concerns the use and settlement architecture of proceeds, not a suspension of ordinary fiscal and administrative discipline.

For the purposes intended here, the word bond has two analytically separate uses. It can name a legal debt instrument, and it can also name a tie, pledge, obligation, or relationship. A social bond in ordinary language is not institutionally equivalent to a security governed by capital-market rules. This distinction matters because regenerative-bond design depends on connecting two layers without confusing them: the formal layer of issuer duties, investor claims, disclosure, custody, and remedies; and the relational layer of local authority, witness, curation, value judgment, limits, repair, and exit.

The principles of green and social bonds improve use-of-proceeds discipline through eligible categories, project selection, proceeds management, and reporting \citep{icma2025gbp,icma2025sbp}. Outcome-oriented instruments add payment or performance logic around measured results, but they need not support local settlement or repair capacity \citep{govuk2026socialoutcomes}. The remaining settlement-architecture gap concerns who gains procurement access, how producers settle obligations, how ecological work becomes payable and repairable, and how communities govern issuance and limits.

\begingroup
\small
\setlength{\tabcolsep}{3pt}
\begin{longtable}{P{0.17\textwidth}P{0.24\textwidth}P{0.22\textwidth}P{0.26\textwidth}}
\caption{Bond instruments and the settlement-architecture gap}
\label{tab:bond-instrument-gap}\\
\toprule
Instrument & Main logic & Strength & Gap addressed by this paper \\
\midrule
\endfirsthead
\toprule
Instrument & Main logic & Strength & Gap addressed by this paper \\
\midrule
\endhead
Conventional bond & Issuer borrows and repays under contract. & Mobilizes capital across time. & Local settlement capacity may remain invisible. \\
Green bond & Proceeds allocated to eligible green projects. & Improves environmental use-of-proceeds discipline. & Procurement, repair, and local production pathways may remain unchanged. \\
Social bond & Proceeds allocated to social projects or target groups. & Improves social use-of-proceeds discipline. & Communities may remain beneficiaries rather than settlement actors. \\
Outcome instrument & Payment or return linked to measured outcomes. & Connects finance to results. & Measurement may still centralize evidence and payment pathways. \\
Regenerative bond & Proceeds allocated to locally governed settlement systems designed to strengthen settlement capacity across locally specified productive, ecological, care, mutual-aid, and repair commitments. & Targets settlement architecture itself. & Requires strong governance, legal review, empirical calibration, and stress testing. \\
\bottomrule
\end{longtable}
\endgroup

\section{Related Literatures}
The argument draws on literatures that are often kept separate. Layered-claim monetary theory treats money as an issuer liability embedded in a hierarchy of settlement and convertibility \citep{ingham2004,minsky1986,mehrling2011}. Commons and reciprocity research emphasizes governance, monitoring, social obligation, and the irreducibility of social meaning to price alone \citep{ostrom1990,mauss2002,polanyi1944,graeber2011}. Rotating-savings and rotating-labor literatures show that mutual-aid institutions can coordinate allocation and obligation outside ordinary bank credit \citep{ardener1964comparative,besley1993economics,anderson2002economics,messerschmidt1981nogar,wang2019rola}.

Market-infrastructure and clearing literature explains why settlement design, netting, margins, and liquidity-saving mechanisms matter for systemic outcomes \citep{brunnermeier2009,duffie2011,cpmiiosco2012}. Netting reduces gross settlement needs by offsetting reciprocal claims before final settlement. Financial-network models warn that connectivity can improve resilience under small shocks while amplifying stress under larger shocks \citep{acemoglu2015systemic}. Obligation-clearing and mutual-credit work shows how network structure can reduce liquidity needs when obligations can be netted or settled through complementary credit \citep{fleischman2020liquidity}. Commitment Pooling literature supplies the operational language of commitments, pooling, curation, valuation, limitation, exchange, and stewardship \citep{ruddick2023letters,ruddick2026proto}. A pool is used here as a governed rule system that can admit listed assets or commitments, apply values and limits, hold inventory or claims, and execute exchange or settlement; the term does not imply legal escrow status.

The empirical strategy is also constrained by recent causal-inference work on network interference. Pool-level liquidity injections can affect other pools through shared redeemable commitments, routing paths, and settlement demand; simple treated-versus-control comparisons are therefore inadequate without an exposure model or an explicit experimental design \citep{leung2022interference,forastiere2024continuous}. The simulation strategy follows agent-based model calibration practice by treating empirical motifs as calibration targets and by separating within-sample fit from stronger holdout or out-of-sample validation \citep{barde2017abm}.

This paper connects these literatures through operational settlement architecture.

\section{Definitions and Boundary Conditions}
\subsection{Regeneration}
Regeneration is defined operationally as an increase in the capacity of ecological, social, cultural, economic, and governance systems to maintain, repair, and renew valued functions over time \citep{fullerton2015regenerative}. The framework treats these claims separately. A transaction trace may show circulation, but it cannot by itself show cultural continuity, consent, ecological regeneration, or welfare improvement.

\begingroup
\small
\setlength{\tabcolsep}{3pt}
\begin{longtable}{P{0.18\textwidth}P{0.38\textwidth}P{0.32\textwidth}}
\caption{Operational dimensions of regeneration}
\label{tab:regeneration-dimensions}\\
\toprule
Dimension & Capacity to be renewed & Evidence test \\
\midrule
\endfirsthead
\toprule
Dimension & Capacity to be renewed & Evidence test \\
\midrule
\endhead
Ecological & Soil, water, biodiversity, food systems, forests, and climate resilience. & Verified ecological reports and local review, not transaction volume alone. \\
Social & Trust, mutual-aid, participation, conflict repair, and collective-action capacity. & Preservation of local curation, repair, and reciprocal settlement motifs. \\
Cultural & Language, memory, ritual, place-based knowledge, and locally meaningful institutions. & Local-language review, practitioner validation, continuity of local terms, and safeguards against cultural branding without authority. \\
Economic & Local production, procurement retention, settlement capacity, livelihood resilience, and reduced leakage. & Settlement velocity, issuer-return proxies, local procurement retention, and leakage diagnostics. \\
Governance & Capacity to curate, value, limit, settle, audit, repair, pause, and exit. & Governance logs, local vetoes, pause rights, appeals, and anti-capture stress tests. \\
\bottomrule
\end{longtable}
\endgroup

\subsection{ROLA-like and ROSCA-like motifs}
Mutual-aid refers to locally recognized practices through which participants support one another through labor, care, goods, services, land work, emergency response, ceremonial work, and general commons tasks \citep{mauss2002,polanyi1944,sahlins1972,kropotkin1902,ostrom1990}. Rotating labor associations, or ROLA-like systems, rotate labor, care, goods, services, or practical support \citep{messerschmidt1981nogar,wang2019rola}. Rotating savings and credit associations, or ROSCAs, rotate monetary or near-money contributions through a pot allocated by rotation, lottery, auction, need, or another rule \citep{ardener1964comparative,geertz1962middle,besley1993economics,anderson2002economics,bouman1995roscas}.

A commitment pool can represent both motifs, but it makes different things movable. In a ROLA-like use, the pool holds and routes redeemable commitments for goods, services, labor, care, ecological work, or other obligations; the relevant motif is multi-asset and often multilateral. In a ROSCA-like use, the pool coordinates stable or cash-like deposits, borrowing rights, repayment obligations, and rotating access to liquidity; the relevant motif is pooled credit and can remain closer to bilateral lender-borrower repayment. Real pools can contain both motifs. The Sarafu Network implementation studied here is analytically useful because redeemable commitments and stable-credit flows interact in the same commitment-pool architecture.

The labels ROLA-like and ROSCA-like are analytical proxies for observed obligation patterns, not direct cultural classifications. They are grounded in the comparative literatures cited above and in the practitioner record used in the Sarafu case, but they do not assert equivalence with every locally named institution. Operationally, ROLA-like evidence is measured through commitment-to-commitment exchange, redeemable-commitment source activity, commitment-issuer return, redemption, routing, clearing, and repair. Repair means a governed response after non-fulfillment, stale obligation, failed redemption, or route failure, such as relisting, limit adjustment, redemption support, guarantor action, write-down, or dispute handling. ROSCA-like evidence is measured through deposits, credit limits, stable borrowing, repayment, mature borrow-proxy closure, and stable recovery. Here mature means that the event has passed the pre-specified observation window used by the audit, so later commitment-issuer return or burn support could in principle be observed.

\subsection{Core definitions}
\begin{definition}[Redeemable commitment]
A redeemable commitment is a witnessed or recorded promise by an issuer to provide a good, service, labor contribution, cash-like redemption, ecological task, or other specified performance under stated terms. It may circulate as an obligation or claim if its terms and law permit; it is not assumed here to be a securities-law debt instrument, and its legal status, transferability, and enforcement depend on its terms and jurisdiction. In the Sarafu Network implementation studied below, such commitments are represented as vouchers; after this point, the paper uses voucher when referring to that implementation or to its Monte Carlo analogue.
\end{definition}

\begin{definition}[Stable-value settlement asset]
A stable-value settlement asset is a cash-like or stable-denominated asset used for settlement in the data or simulation. The term is a modeling and taxonomy category; it does not imply legal tender status, deposit insurance, bank-money status, accounting cash-equivalent treatment, or frictionless convertibility unless those properties are explicitly established in the relevant jurisdiction and instrument terms.
\end{definition}

\begin{definition}[Commitment pool]
A commitment pool is a locally stewarded institution or contract suite that implements four functions: curation of admissible commitments, valuation through a value index, limitation of exposure or flow, and exchange or settlement \citep{cicStackDocs,erc20PoolRepo,clcProtocolOverview}.
\end{definition}

\begin{definition}[Regenerative bond]
A regenerative bond is a formal debt instrument whose disclosed use-of-proceeds and governance rules allocate proceeds to locally governed settlement systems designed to strengthen settlement capacity across locally specified productive, ecological, care, mutual-aid, and repair commitments. This capacity is implemented through commitment pools and related settlement architecture, including route graphs, procurement lanes, reserves, reporting, and repair pathways where specified. The formal issuer remains responsible for all scheduled payment, disclosure, covenant, and remedy obligations under the bond terms, including principal and coupon or equivalent return; local pools, households, stewards, and redeemable commitments are not bondholder collateral unless explicitly and legally structured as such.
\end{definition}

This definition identifies a governed use-of-proceeds and settlement architecture; it does not by itself assert demonstrated welfare, ecological, cultural, or legal outcomes.

\begin{definition}[Formal issuer]
A formal issuer is the government, municipality, development institution, regulated vehicle, cooperative federation, or other legally authorized entity that owes scheduled principal, coupon or equivalent return, disclosure, covenant, and remedy obligations to bondholders under the bond terms. Local pools and households are not formal issuers unless explicitly and legally structured as such.
\end{definition}

\begin{definition}[Eligible service resources]
Eligible service resources are cash, stable-value assets, reserves, recovered pool stable, or contractually available revenues that the issuer may legally use for scheduled bond servicing. Local in-kind commitments and ecological work may create public value, but they are not eligible debt-service resources unless converted through lawful and explicit mechanisms. Such mechanisms are addressed in Section \ref{sec:governance-law-non-extraction}.
\end{definition}

The eligible-service-resource definition excludes generic pool turnover, symbolic accounting, or in-kind ecological work from debt-service resources unless a lawful conversion pathway is specified.

A bond is not described as regenerative in this framework unless five minimum boundary conditions hold:
\begin{enumerate}[leftmargin=1.5em]
\item the formal issuer remains responsible for scheduled payment, disclosure, covenant, and remedy obligations to bondholders under the bond terms;
\item local pools retain meaningful authority over curation, valuation, limits, repair, and exit;
\item proceeds target real settlement capacity across locally specified productive, ecological, care, mutual-aid, and repair commitments, rather than generic turnover or symbolic accounting;
\item reporting distinguishes cash payment, fulfillment, clearing, repair, and loss recognition; and
\item the system includes safeguards against coercive repayment, speculative claims, cultural flattening, data extraction, and hidden household guarantees.
\end{enumerate}

\section{Commitment Pools and the Institutional Membrane}
Commitment Pooling Protocol (CPP) is used in this paper as an operational grammar for settlement design \citep{ruddick2023letters,ruddick2026proto,ruddick2026routegraphs}. CPP is used here as a minimal resource-coordination protocol rather than as a complete theory of welfare, culture, law, or value. The protocol asks only the operational questions that any pool-mediated resource system must answer before it can coordinate commitments: what may be admitted, how admitted commitments are valued, how much exposure or flow is allowed, and how accepted commitments move, settle, or enter repair. This makes CPP useful for the present paper because regenerative-bond proceeds are not simply disbursed into projects; they enter a local settlement environment where producers, stewards, pools, public agencies, stable-value assets, and redeemable commitments interact under rules.

This minimality matters. A generic ledger can record balances without saying which commitments are admissible, who may change limits, what happens when settlement fails, or which flows can lawfully support formal debt service. Conversely, a full ethnographic or legal account can describe legitimacy and authority without providing a functional interface for routing, stress testing, or issuer-service accounting. CPP sits between these two levels. It supplies a small set of interfaces for resource coordination while leaving legal status, cultural meaning, welfare impact, and ecological verification to separate evidence and governance layers. In the empirical sections, Sarafu Network is treated as one implementation of this grammar. In the Monte Carlo sections, the same grammar becomes the state and routing substrate used to test settlement-capacity frontiers under explicit assumptions.

Curation asks what commitments may enter. Valuation asks how those commitments are priced or indexed. Limitation asks what prevents over-issuance, over-exposure, coercion, and harm. Exchange asks how commitments move, settle, route, generate receipts, trigger fees, and enter repair. For example, a producer voucher accepted by a pool may be exchanged for stable value only if it is listed, valued, within the exposure limit, and supported by pool inventory; later repayment, redemption, or circulation can remove that voucher from pool exposure without making households liable to bondholders.

This grammar differs from a static bilateral loan. A bilateral lender-borrower obligation can stagnate when the borrower cannot repay the lender directly, even if the borrower still has useful productive capacity or redeemable commitments that others would accept. A pool-held voucher can move: another participant can buy it with stable value, borrow it with another voucher, route it through overlapping pools, or return it to its commitment issuer through fulfillment, redemption, or repair. Such movement does not erase the formal bond issuer's liability, and it does not by itself create bond-service cash. However, it can reduce pool-level exposure, create additional settlement paths, and prevent local obligations from remaining frozen within a single relationship.

The reference architecture separates seven accounting and governance functions: investor claim, issuer obligation, settlement facility, eligible pools, local commitment network, eligible service resources, and safeguards. Investors hold claims on the formal issuer, not on local pools or households. Bond proceeds enter a regenerative settlement facility and then eligible pools. Only explicit eligible service resources, such as recovered stable value, cash-eligible fees, and issuer reserves, may support scheduled bond service.

\begin{figure}[H]
\centering
\includegraphics[width=\textwidth]{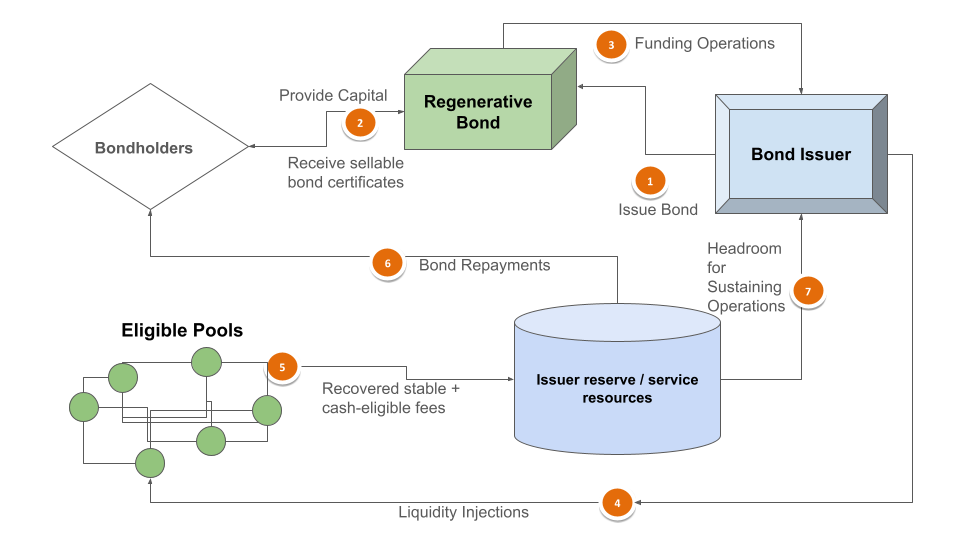}
\caption{Conceptual regenerative-bond funding and service architecture. 
(1) The formal issuer authorizes and issues the regenerative bond. 
(2) Bondholders provide capital and receive transferable bond certificates or claims against the issuer. 
(3) Bond proceeds support the issuer's funding operations under the disclosed bond program. 
(4) The issuer deploys liquidity into eligible pools. 
(5) Eligible pools return recovered stable value and cash-eligible exchange fees to issuer reserve and service resources under the stated allocation rules. 
(6) Scheduled bond repayments are made to bondholders from issuer service resources. 
(7) Resources above scheduled service constitute candidate headroom for sustaining operations, monitoring, certification, reserves, and risk costs. 
The figure is an accounting and governance schematic: bondholders hold claims on the issuer, not on local pools, households, producer vouchers, or mutual-aid obligations.}
\label{fig:regenerative-bond-flow}
\end{figure}

Figure~\ref{fig:regenerative-bond-flow} summarizes the accounting boundary before Table~\ref{tab:institutional-architecture} describes the institutional responsibilities.

\FloatBarrier

\begingroup
\small
\setlength{\tabcolsep}{3pt}
\begin{longtable}{P{0.19\textwidth}P{0.35\textwidth}P{0.34\textwidth}}
\caption{Regenerative-bond institutional architecture}
\label{tab:institutional-architecture}\\
\toprule
Layer & Primary obligations & Not to be confused with \\
\midrule
\endfirsthead
\toprule
Layer & Primary obligations & Not to be confused with \\
\midrule
\endhead
Formal issuer
& Coupon, principal, disclosures, covenants, fiscal-risk management, and legal remedies owed to investors.
& Local household commitments or village-level relational obligations. \\
\addlinespace
Capital-market interface
& Investor onboarding, listing or placement, custody, disclosure, transfer rules, trustee or fiscal-agent functions.
& Local stewardship or cultural authority. \\
\addlinespace
Regenerative settlement facility
& Reserve custody, allocation rules, audit, reporting, insurance or guarantee administration, liquidity mandates, and published waterfall.
& Investor ownership of local governance or automatic cash generation from local commitments. \\
\addlinespace
Reserve and guarantee layer
& Insurance reserves, liquidity buffers, guarantor rules, claim triggers, payout caps, and recovery rules.
& Unlimited protection or implicit household guarantee. \\
\addlinespace
Eligible local commitment pool
& Curation, valuation, limits, receipts, settlement paths, repair rules, pause rights, and local accountability.
& Formal bond guarantor or unrestricted borrower. \\
\addlinespace
Commitment network
& Households, producers, cooperatives, service providers, public agencies, institutions, and buyers issue, accept, fulfill, redeem, or repair commitments.
& A passive beneficiary population or collateral pool. \\
\addlinespace
Reporting and verification
& Receipts, governance logs, ecological indicators, social indicators, pool health, risk events, disputes, and repair outcomes.
& Proof of regeneration from transaction volume alone. \\
\bottomrule
\end{longtable}
\endgroup

\subsection{Accounting boundary}
For commitment issuer $a$, pool $p$, and voucher class $j$, let $C_{apj}$ denote the accepted credit or flow limit for that commitment issuer's commitments, valued in a common index. Let $D_{apj}$ denote outstanding unsettled commitments. The remaining accepted settlement headroom is:
\[
B_{apj} = C_{apj} - D_{apj}.
\]

This is a settlement-capacity identity, not a corporate-equity identity: more issued vouchers do not imply more capacity unless pools accept, value, and limit them. Settlement flows enter separately below through $S$, the obligation-closing flow over a measurement interval.

Bond debt service is an issuer obligation:
\[
DS_t = \mathrm{coupon}_t + \mathrm{principal}_t.
\]
Eligible issuer service resources are:
\[
E^B_t = H_t + \alpha_B \Phi_t + R^{pay}_t,
\]
where $E^B_t$ denotes bond-service-eligible resources, the superscript $B$ denotes bond-service eligibility, $H_t$ is eligible recovered stable resources from pools, $\Phi_t$ is cash-eligible fee flow, $\alpha_B \in [0,1]$ is a published allocation share, and $R^{pay}_t$ is legally available reserve draw for scheduled service if the scenario includes one. Required or prefunded reserves are not surplus merely because they exist; they count toward payment only when instrument terms permit a draw. The empirical implementation restricts bond-service eligibility to explicitly recovered stable resources and excludes generic pool inventory turnover from issuer service resources. The reported frontier uses zero issuer reserve share. The debt-service coverage ratio is:
\[
\kappa_t = \frac{E^B_t}{DS_t}
\quad \text{for } DS_t > 0.
\]
Arrears are:
\[
A^B_t = \max(0, DS_t - E^B_t).
\]
A scheduled-service shortfall occurs when:
\[
\mathbf{1}_{\mathrm{shortfall},t} =
\begin{cases}
1, & E^B_t < DS_t,\\
0, & E^B_t \geq DS_t,
\end{cases}
\]
before any instrument-specific cure, deferral, capitalization, grace-period, or restructuring terms are applied. Default remains a legal status defined by the instrument rather than by this diagnostic alone.
Gross issuer service headroom before operating, monitoring, legal, governance, audit, and risk-capital costs is:
\[
X^B_t = \max(0, H_t + \alpha_B \Phi_t + R^{rel}_t - DS_t).
\]
Here $R^{rel}_t$ is reserve value released for unrestricted issuer use after scheduled service and required reserve targets are met; otherwise it is zero. This headroom variable is not a profit measure, reserve requirement, or investor claim. It is a diagnostic for whether scheduled service leaves candidate resources for operating, monitoring, governance, legal, audit, and risk-capital costs.

For the reported frontier guardrail, the corresponding headroom statistic is expressed as a coverage multiple:
\[
\eta^B_t = \frac{H_t + \alpha_B \Phi_t + R^{rel}_t}{DS_t}
\quad \text{for } DS_t > 0.
\]
The guardrail $\eta^B_{p50} \geq 1.25$ means that median modeled service-and-release resources are at least 25 percent above scheduled service before full operating and risk costs.

A related velocity-of-settlement lens is useful but narrower than ordinary transaction volume. For a voucher or obligation class $j$, let $D_{jt}$ be unsettled obligation stock and $S_{jt}$ be obligation-closing settlement flow in period $t$. A settlement-velocity diagnostic is $\nu_{jt}=S_{jt}/D_{jt}$, with appropriate period averaging when $D_{jt}$ changes over the interval. Federated routing can raise $\nu$ by helping obligations find acceptable redemption, buyback, netting, or repair paths. Fee volume, however, is generated by executed routed value and may include movement that does not close obligations. The paper therefore treats fee cash as eligible only after asset eligibility, conversion, and lockbox rules are applied. Asset eligibility means legal availability for bond service; conversion means movement into an eligible settlement asset; lockbox rules mean disclosed segregation and allocation rules for issuer service. The paper does not infer debt-service capacity from gross velocity or gross turnover alone.

The model is governed by five design invariants: formal bondholder claims must remain separate from local relational obligations; local stewards must retain curation authority; advances must remain bounded by valuation, evidence, limits, and haircuts; repair pathways must have explicit triggers and state updates; and debt service must not depend primarily on scarce cash extraction from vulnerable households. A haircut is a discount applied to the accepted value of a commitment or asset for risk, liquidity, or uncertainty. The full invariant table is reported in the online supplement.

\section{Sarafu Network as Component-Level Evidence}
Sarafu Network is treated as a component-level implementation and calibration case, not as a historical regenerative bond \citep{ruddick2026sarafuDataset}. The retained cohort is restricted to Kenyan-shilling-denominated community pools with enough observed activity to support calibration. It contains 73 pools, 996 unique producer voucher wallets in 1,247 accepted voucher member slots, and 462 external non-producer stable-side users. Pools are rule-governed exchange venues. Producer and consumer holdings in the empirical topology are private route endpoints rather than intermediate swap venues; formal issuer and bondholder states are added only in the counterfactual bond simulation. Sarafu Network implemented key parts of the commitment-pool architecture: vouchers represent redeemable commitments; pools curate and exchange them; limits and exchange-rate records make acceptance capacity observable; stable or cash-like assets interact with vouchers; and field reports provide a separate ecological and welfare evidence layer.

The evidence consists of linked motifs connecting social obligation, pool exchange, stable liquidity, routing, and report-linked outcomes. The component evidence is reported below, while the calibration map appears within Table \ref{tab:sarafu-moments} in the discussion on Monte Carlo calibration.

\subsection{Pool activity and report-linked evidence}
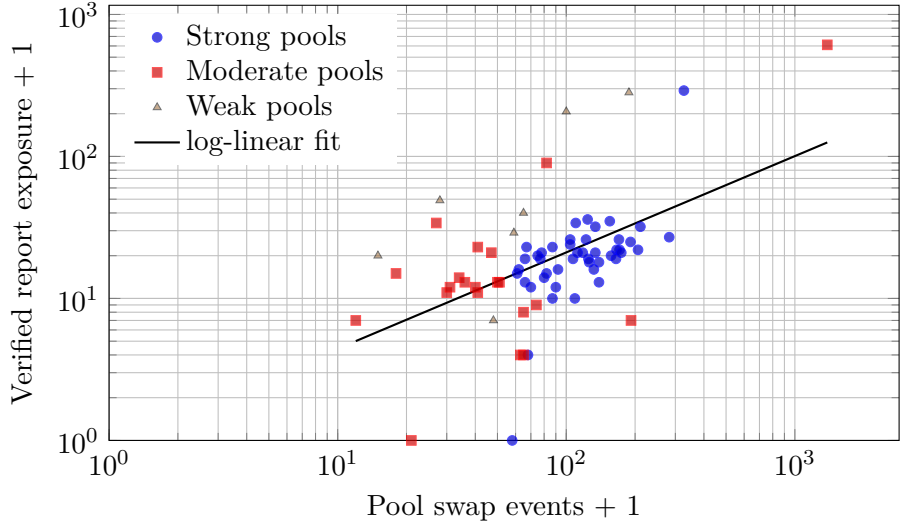
\begin{figure}[H]
\centering
\begin{tikzpicture}
\begin{axis}[
  width=0.95\textwidth,
  height=0.58\textwidth,
  xmode=log,
  ymode=log,
  xmin=1,
  ymin=1,
  xlabel={Pool swap events + 1},
  ylabel={Verified report exposure + 1},
  grid=both,
  legend style={at={(0.02,0.98)}, anchor=north west, draw=none, fill=white},
  legend cell align={left},
]
\addplot+[only marks, mark=*, mark size=1.8pt, blue, opacity=0.70]
  table[x=swap_events_plus1,y=verified_report_exposure_plus1,col sep=comma]
  {analysis/pool_report_activity_strong.csv};
\addplot+[only marks, mark=square*, mark size=1.8pt, red, opacity=0.68]
  table[x=swap_events_plus1,y=verified_report_exposure_plus1,col sep=comma]
  {analysis/pool_report_activity_moderate.csv};
\addplot+[only marks, mark=triangle*, mark size=1.7pt, black, opacity=0.55]
  table[x=swap_events_plus1,y=verified_report_exposure_plus1,col sep=comma]
  {analysis/pool_report_activity_weak.csv};
\addplot+[no marks, thick, black]
  table[x=swap_events_plus1,y=verified_report_exposure_plus1,col sep=comma]
  {analysis/pool_activity_verified_fit.csv};
\legend{Strong pools, Moderate pools, Weak pools, log-linear fit}
\end{axis}
\end{tikzpicture}
\caption{Pool activity and verified impact-report exposure. Each point is a privacy-safe pool. The fitted line summarizes a descriptive log-log association. Strong, moderate, and weak are review tiers based on recency and continuity of observed exchange, not final truth labels.}
\label{fig:pool-activity-reporting}
\end{figure}

Pool-level evidence is consistent with active commitment pools also being sites where report-linked impact infrastructure is more visible. A privacy-safe cross-section links pool ranking data to field-report exposure through voucher-pool membership. The report table contains 1,383 unique verified reports; because a report can link to more than one pool through voucher memberships, the full-count pool-linked verified-report exposure across included pools is 2,663. Across the 73 included community pools, 71 have nonzero verified-report exposure. Figure~\ref{fig:pool-activity-reporting} uses log-scaled axes because both count variables are highly skewed; the fitted line is descriptive only. On plus-one log-transformed values, Pearson correlation is 0.515 and Spearman rank correlation is 0.460. This association is useful for calibration and deployment-site selection; it is not causal proof that pool activity produces welfare or ecological impact.

The strong, moderate, and weak pool tiers are review categories used for calibration and diagnostics. Strong candidates have a last swap within 60 days of the dataset's latest observed pool swap and at least four swap-active weeks in the trailing 90-day window, with additional activity and reporting-readiness filters. Moderate candidates have a last swap within 90 days and at least two swap-active weeks in the trailing 90-day window. Weak candidates do not meet those thresholds. These tiers are not welfare, ecological, or certification labels.

The report-linked layer also provides activity-category composition. In Table~\ref{tab:report-activity-categories}, $\hat{y}_{100}$ denotes the fitted verified-report exposure at 100 pool swap events under the descriptive log-log model, and elasticity is the fitted log-log slope for that reported activity category. Table \ref{tab:report-activity-categories} reports the largest public activity categories by verified report exposure. These counts are report-exposure diagnostics linked through vouchers, not unique beneficiary counts or estimates of impact. Verified reports provide evidence infrastructure that may support future certification after verification protocol review, local review, and domain validation.

\begin{table}[H]
\centering
\caption{Verified report exposure by reported activity category}
\label{tab:report-activity-categories}
\begingroup
\small
\setlength{\tabcolsep}{3pt}
\begin{tabular}{lrrrr}
\toprule
Reported activity & Verified exposure & Share & $\hat{y}_{100}$ & Elasticity \\
\midrule
Regenerative Agriculture & 599 & 22.5\% & 2.92 & 0.88 \\
Cleaning & 372 & 14.0\% & 1.97 & 0.39 \\
Construction & 303 & 11.4\% & 1.94 & 0.61 \\
Training & 220 & 8.3\% & 1.70 & 0.34 \\
Housekeeping & 200 & 7.5\% & 1.02 & 0.29 \\
Permaculture & 115 & 4.3\% & 0.68 & 0.36 \\
Lending & 104 & 3.9\% & 0.65 & 0.26 \\
Education & 101 & 3.8\% & 0.81 & 0.52 \\
\bottomrule
\end{tabular}
\endgroup

\end{table}
\FloatBarrier

\subsection{Same-token return balance and cash/stable recovery}
In the Sarafu Network data, a token is a digital ledger entry representing either a stable/cash-like settlement asset, a redeemable voucher, or an internal voucher category. Cash/stable tokens are cash-like or stable-denominated settlement entries in the taxonomy. Redeemable vouchers represent producer or community commitments. Internal vouchers are system or pool-internal entries used for accounting, testing, or internal settlement categories.

For pool repayment mechanics, the clearest diagnostic is a pool-token flow balance. The FIFO audit treats a swap out of the pool as open exposure and matches later swaps of the same token back into that pool, capped at the original outflow. The calibration cohort identifies 914 pool-token pairs; matched same-token swap-in covers 38,605.51 of 89,210.79 KES-denominated swap-out value, or 43.3 percent. Redeemable vouchers show 88.2 percent same-token return coverage, internal vouchers 64.5 percent, and cash or stable tokens 23.8 percent. The supplement reports the aggregate audit support.

The reported cash/stable return coverage uses a broader recovery interpretation: stable-to-voucher swap-in value into included pools relative to voucher-to-stable advances out of those same pools. The estimated coverage is 26.2 percent. Broader replenishment is treated as a liquidity-inflow diagnostic rather than repayment or issuer service capacity.

This distinction is essential: replenishment may indicate liquidity entering a pool, but only eligible recovered stable tokens can support issuer service in the bond model.

The producer-voucher integrity audit summarized in the supplement checks the wallet-to-pool accounting boundary and confirms that the retained cohort should be interpreted as pool-level exposure and settlement evidence rather than as a legal debt ledger. The supplement also reports mint-event, pool-deposit, credit-limit, swap-in, and flow-balance details.

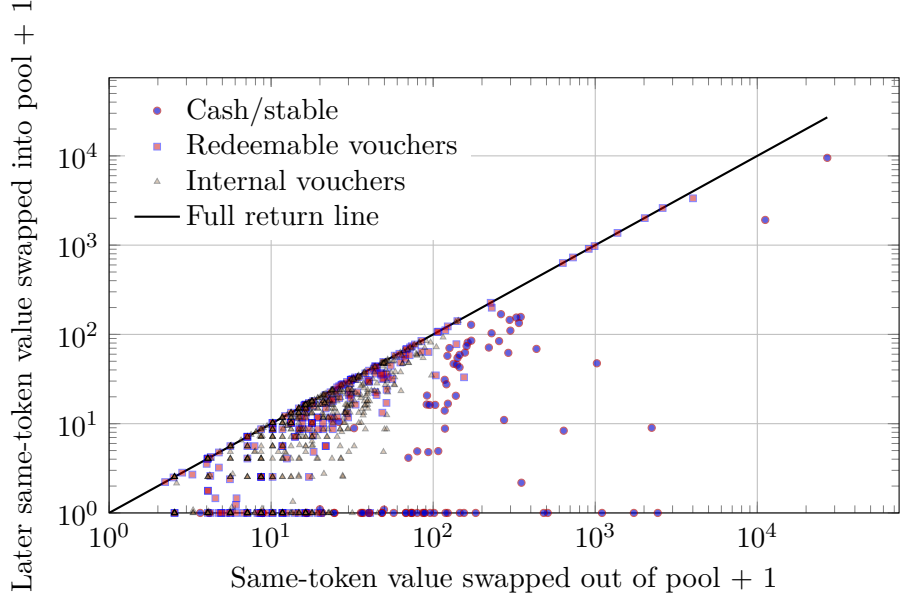
\begin{figure}[H]
\centering
\begin{tikzpicture}
\begin{axis}[
  width=0.95\textwidth,
  height=0.58\textwidth,
  xmode=log,
  ymode=log,
  xmin=1,
  ymin=1,
  xlabel={Same-token value swapped out of pool + 1},
  ylabel={Later same-token value swapped into pool + 1},
  grid=major,
  legend style={at={(0.02,0.98)}, anchor=north west, draw=none, fill=white},
  legend cell align={left},
]
\addplot+[only marks, mark=*, mark size=1.35pt, red!70!black, opacity=0.62]
  table[x=out_value_plus1,y=matched_later_in_value_plus1,col sep=comma]
  {analysis/same_token_return_cash.csv};
\addplot+[only marks, mark=square*, mark size=1.25pt, blue, opacity=0.48]
  table[x=out_value_plus1,y=matched_later_in_value_plus1,col sep=comma]
  {analysis/same_token_return_redeemable_voucher.csv};
\addplot+[only marks, mark=triangle*, mark size=1.20pt, black, opacity=0.36]
  table[x=out_value_plus1,y=matched_later_in_value_plus1,col sep=comma]
  {analysis/same_token_return_internal_voucher.csv};
\addplot+[no marks, thick, black]
  table[x=out_value_plus1,y=matched_later_in_value_plus1,col sep=comma]
  {analysis/same_token_return_identity.csv};
\legend{Cash/stable, Redeemable vouchers, Internal vouchers, Full return line}
\end{axis}
\end{tikzpicture}
\caption{Same-token return balance by pool-token pair. Points on the diagonal are fully returned at pool level; points below the diagonal are unresolved or only partially returned. The figure provides direct settlement-flow evidence without identifying the off-chain reason for return or the individual borrower.}
\label{fig:same-token-return-balance}
\end{figure}
\FloatBarrier

\section{Monte Carlo Model and Calibration Fit}
The Monte Carlo model uses Sarafu Network as calibration evidence for stress testing. Its purpose is to explore the mechanism under stated assumptions: which assumptions matter, where the mechanism breaks, how sensitive outcomes are to parameters, and which variables must be measured before deployment.

The public simulator supplies the computational starting point \citep{cosmoLocalCreditSim}. In the reported topology, the empirical community pools are modeled as traversable liquidity and settlement pools, a simulator class for rule-governed liquidity and settlement venues rather than a legal lending classification. Producers and consumers have private wallets. They can start or receive a route, but ordinary routing and clearing traverse pools only. The bond layer adds formal issuer state, bond principal, coupon schedule, eligible pool deployment, producer-debt maturity, pool recovery, scheduled-service accounting, and issuer headroom. Detailed agent actions, routing, clearing, redemption timing, and sensitivity design are reported in the supplement.

The calibration target is the 73-pool Kenya commitment pool cohort. Moments are empirical summary statistics used as calibration targets or diagnostics. Table \ref{tab:sarafu-moments} is therefore a calibration map, not a standalone proof of impact. The generated calibration artifacts report zero binding failures under the pre-specified within-sample checks across pool count, swap activity, report exposure, asset return, cash return, voucher return, borrow-proxy closure, tier input scale, and ledger-observed settlement motifs, using 100 Monte Carlo runs over 260 weekly ticks. Because the paper does not report a reserved holdout period, pool set, or token class, the simulation is calibrated stress testing. Independent replication is especially important because author practitioner involvement, token classification, calibration choices, and simulation outputs affect policy or investment interpretation.

\begin{table}[H]
\centering
\caption{Sarafu moments used for calibration and interpretation}
\label{tab:sarafu-moments}
\small
\setlength{\tabcolsep}{3pt}
\begin{tabularx}{\textwidth}{P{0.34\textwidth}P{0.20\textwidth}P{0.36\textwidth}}
\toprule
Motif & Main observed moment & Interpretation \\
\midrule
Included community pools & 73 pools & Kenya KES community-pool cohort. \\
Accepted producer voucher members & 996 wallets / 1,247 slots & Producer vouchers may be accepted in multiple pools. \\
External non-producer stable users & 462 wallets & Consumer-wallet count used for topology, distinct from producer wallets. \\
Voucher-to-voucher swaps & 5,103 events & ROLA-like circulation proxy. \\
Trailing 90-day voucher-to-voucher share & 81.4\% & Recent high voucher-circulation regime. \\
Mature borrow-proxy event support & 56.5\% & ROSCA-like closure proxy over 90-day mature events. \\
Mature borrow-proxy value support & 64.9\% & Value-weighted stable-credit closure. \\
Same-token return coverage & 43.3\% & Pool-level return balance. \\
Redeemable voucher return coverage & 88.2\% & Voucher return stronger than cash/stable return. \\
Cash/stable return coverage & 26.2\% & Stable-to-voucher recovery relative to voucher-to-stable stable advances. \\
All cash/stable replenishment coverage & 101.6\% & Liquidity-inflow diagnostic, not repayment. \\
Stable pool replenishment rate & 24.64 / pool-month & Stable explicit deposits plus direct non-swap transfers into included pools. \\
Voucher pool replenishment rate & 9.18 / pool-month & Voucher explicit deposits plus direct non-swap transfers into included pools. \\
Largest pool-token graph component & 67 pools / 938 vouchers & Settlement-topology potential. \\
Verified reports / pool-linked exposure & 1,383 / 2,663 & Unique verified reports and full-count pool-linked exposure; outcome-evidence infrastructure, not impact proof. \\
\bottomrule
\end{tabularx}
\end{table}

The simulation reports the metrics needed for the stated frontier claim: voucher-to-voucher value, stable recovery, fee-service cash, eligible recovered stable resources, scheduled-payment coverage, issuer headroom, leakage, concentration, and cash-stress diagnostics. These metrics separate three questions: whether bondholders receive scheduled payment, whether the issuer has modeled headroom beyond scheduled service, and whether voucher circulation and stable-credit closure are preserved relative to matched no-bond baselines.

Several assumptions remain outside the reported result. Public procurement retention, governance overrides, explicit operating-cost schedules, and insurance-claim dynamics are model extensions requiring separate calibration. The route pass rate is a model stress screen, not a directly observed Sarafu route-success statistic. Empirical flows are KES-denominated, while simulated stable assets are modeled stable-value units; stable/cash comparisons are therefore dependency and service-resource diagnostics, not claims of frictionless legal convertibility. The supplement reports the additional settlement-timing assumptions and monitoring priorities.

\section{Governance, Law, and Non-Extraction Safeguards}
\label{sec:governance-law-non-extraction}
Governance is a boundary condition for the model, not a compliance afterthought. A regenerative bond can fail through over-valuation, over-borrowing, cultural flattening, central capture, mis-selling, legal confusion, data extraction, technical exclusion, or debt-service pressure.

Jurisdiction-specific legal review is required for the formal bond, issuer, settlement facility, pool lending or advances, vouchers, fees, secondary trading, custody, tax, consumer protection, data governance, and payment-system classification. Public-sector implementations also require fiscal-risk review \citep{worldbankimf2014debt}. The design must not move public liabilities off balance sheet or obscure who bears default risk.

Securities disclosure should state explicitly that investors have claims against the formal issuer or regulated vehicle, not against local pools, households, vouchers, elders, stewards, or mutual-aid practices. Any reserve, guarantee, fee waterfall, or service-cash allocation should be disclosed as a bounded source of issuer support, not as a direct investor claim on community assets.

Local governance rights include curation, valuation, limits, repair, evidence, pause, exit, appeal, cultural review, and data review. A pool is locally governed only if local stewards have real authority over admissible commitments, valuation references, exposure limits, repair paths, and exit.

These safeguards define the boundary between a regenerative-bond settlement architecture and ordinary debt expansion.

Operational safeguards include exposure caps, haircuts, evidence requirements, plain-language disclosures, no guaranteed-yield language, clear non-fulfillment triggers, dispute pathways, local vetoes, audit trails, exit rights, and non-smartphone access where feasible. Public anonymization is not sufficient: network reconstruction and inferred creditworthiness can expose communities to extraction, stigma, surveillance, or unfavorable financial profiling. The empirical protocol therefore relies on minimization, privacy-safe aggregation, governance review, and restrictions on using data to rank communities or sell financial intelligence \citep{carroll2020care,smith2012decolonizing}.

\section{Settlement-Capacity Frontier Specification}
The simulation results are a guardrail frontier, not a deployment recommendation. The model asks whether repayable bond liquidity can coexist with local voucher circulation and stable-credit repayment under Sarafu-calibrated assumptions. These are model results under stated assumptions and before modeling the full operating costs, taxes, legal costs, insurance claims, procurement politics, household welfare, ecological verification, and cultural legitimacy.

The reported frontier is the set of coupon and principal combinations tested for whether modeled service coverage and local-circulation safeguards pass. It is initialized as a Sarafu-calibrated ROSCA-like credit-pool substrate with voucher-capable rules, not as a literal Sarafu replay and not as a pure voucher-free ROSCA. Under those rules, each matched no-bond baseline is generated by the runner for comparison with the corresponding positive-principal scenario. Producer wallets and consumer wallets are private route endpoints; routing and clearing venues are the pools. The positive-principal scenarios then ask whether formal bond liquidity protects, degrades, or amplifies ROLA-like voucher circulation while preserving ROSCA-like stable-credit repayment and issuer responsibility for formal debt service.

The reported frontier uses 100 runs for each of 64 principal-coupon scenarios plus matched no-bond baselines; the resulting classifications are read as simulation guardrail outcomes for the specified runner, configuration, and seed policy rather than universal pass probabilities. Principal ratio means gross bond principal divided by modeled eligible backing capacity, the simulator's pre-specified pool-backing denominator; it is not a legal collateral valuation, third-party certification, or guarantee. The schedule uses a 260-week term, 26-week issuer-payment stride, even principal amortization, coupon accrual on start-of-period outstanding principal, zero issuer reserve share, and a remaining-schedule lockbox. High coupon values are stress tests, not proposed financing terms. The pass region preserves scheduled service, ROLA-like V2V circulation, and issuer headroom; failure begins when debt service, exposure limits, stable-seeking behavior, or excessive liquidity capture local settlement attention.

\begin{table}[htbp]
\centering
\caption{Settlement-capacity frontier design}
\label{tab:settlement-capacity-frontier-design}
\scriptsize
\setlength{\tabcolsep}{3pt}
\begin{tabularx}{\textwidth}{P{0.30\textwidth}P{0.58\textwidth}}
\toprule
Dimension & Reported setting \\
\midrule
Network scale & Current calibrated topology with matched no-bond baselines. \\
Coupon targets & 0\%, 4\%, 8\%, 12\%, 15\%, 25\%, 35\%, and 45\%. \\
Principal ratios & 0.05, 0.10, 0.25, 0.50, 0.75, 1.00, 1.50, and 2.00 of modeled eligible backing capacity. \\
Debt-service schedule & 260-week term; 26-week payment stride; even principal amortization; coupon accrues on start-of-period outstanding principal. \\
Runs per cell & 100 Monte Carlo replications per positive-principal cell, with matched no-bond baselines. \\
Grid size & 64 positive-principal cells. \\
Role & Settlement-capacity frontier and reported guardrail result. \\
\bottomrule
\end{tabularx}
\end{table}

\begin{table}[htbp]
\centering
\caption{Frontier guardrails and reported thresholds}
\label{tab:frontier-guardrail-definitions}
\small
\setlength{\tabcolsep}{4pt}
\begin{tabularx}{\textwidth}{P{0.30\textwidth}P{0.25\textwidth}P{0.35\textwidth}}
\toprule
Guardrail & Reported threshold & Interpretation \\
\midrule
Scheduled-service coverage & p05 $\geq 1.0$ & Fifth-percentile modeled eligible service resources cover scheduled coupon and principal due. \\
Voucher-to-voucher (V2V) preservation & $\Delta \mathrm{V2V} \geq -15\%$ & V2V value does not fall by more than the pre-specified material-decline tolerance relative to the matched no-bond baseline. \\
Active-float preservation & $\Delta \mathrm{ActiveFloat} \geq -15\%$ & The modeled stock of active routable producer-voucher commitments remains within the pre-specified tolerance relative to baseline. \\
Route reliability & p05 route success $\geq 85\%$ & Fifth-percentile route success remains above the model settlement-reliability floor. \\
Issuer operating/risk headroom multiple & p50 $\geq 1.25$ & Median modeled service-and-release resources are at least 25 percent above scheduled service before full operating and risk costs. \\
\bottomrule
\end{tabularx}
\end{table}

The reported 64-cell frontier is a guardrail frontier rather than a single maximum bond size. Here p05 denotes the fifth percentile and p50 denotes the median across replications. Scheduled bondholder payment clears in 57 cells. V2V value is preserved within the material-decline threshold in all 64 cells and increases relative to the matched no-bond baseline in 24 cells. Fifth-percentile route success remains above the 85 percent floor in all cells. Median issuer operating/risk headroom remains at least 1.25 in 52 cells; the same 52 cells form the strong-success subset by clearing scheduled payment, V2V and active-float preservation, route reliability, and headroom.

\begin{table}[htbp]
\centering
\caption{Reported 64-cell settlement-capacity frontier summary}
\label{tab:settlement-capacity-frontier-summary}
\scriptsize
\setlength{\tabcolsep}{4pt}
\begin{tabularx}{\textwidth}{P{0.48\textwidth}P{0.18\textwidth}P{0.26\textwidth}}
\toprule
Guardrail or diagnostic & Cells & Interpretation \\
\midrule
Scheduled payment coverage p05 $\geq 1.0$ & 57 / 64 & Formal scheduled bond service clears through most tested cells and begins to fail at higher coupon and higher principal stress. \\
Voucher-to-voucher value preserved within material-decline threshold & 64 / 64 & ROLA-like circulation is preserved under the material-decline guardrail across the tested grid. \\
Strict voucher-to-voucher value increase & 24 / 64 & Some cells amplify V2V value relative to their matched no-bond baselines, but amplification is not required for minimum safety. \\
Route success p05 $\geq 85\%$ & 64 / 64 & Settlement reliability remains above the model route-success floor in every tested cell. \\
Issuer headroom multiple p50 $\geq 1.25$ & 52 / 64 & Most cells retain at least 25 percent median candidate operating and risk margin above scheduled service. \\
Strong settlement-capacity subset & 52 / 64 & Cells that clear scheduled service, preserve V2V and active-float guardrails, and retain at least a 1.25 median issuer headroom multiple. \\
\bottomrule
\end{tabularx}
\end{table}

The strong subset is not an optimal financing region; it is the intersection of scheduled service, V2V and active-float preservation, route reliability, and candidate issuer headroom.

The final grid is intentionally edge-inclusive: it includes high principal and coupon combinations to reveal where guardrails begin to bind. Scheduled-payment coverage becomes the visible boundary at larger principal and coupon combinations, simulated settlement motifs shift across the grid, and the outcome grid separates low-stress pass regions from higher-stress review or failure regions. The static 5x deposit-based borrowing cap is a modeled guardrail limiting producer borrowing relative to deposited or accepted voucher capacity. Cap-bound producer counts are not a central result; borrowing-cap edges are treated as sensitivity diagnostics.

\begin{figure}[htbp]
\centering
\includegraphics[width=0.9\textwidth]{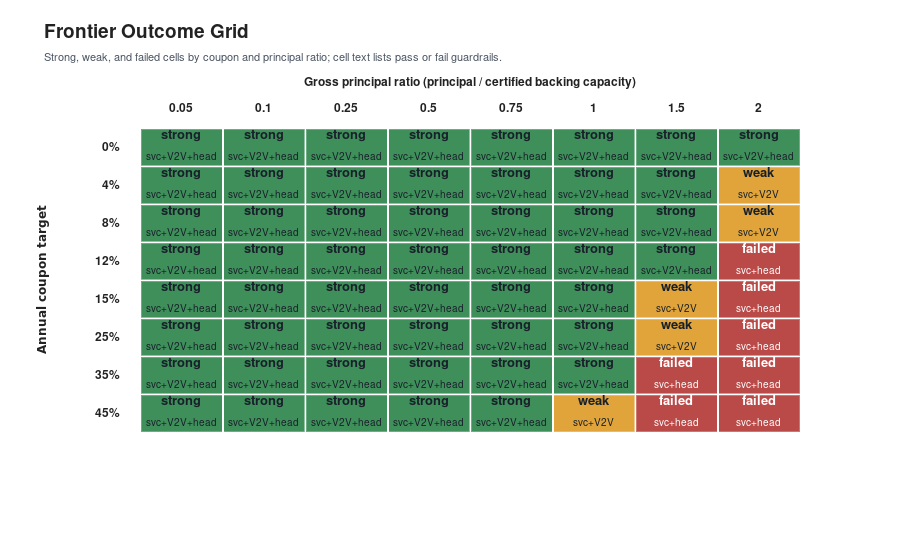}
\caption{Final settlement-capacity grid: frontier outcome grid. Strong cells pass scheduled-service, V2V-preservation, active-float, route-reliability, and issuer-headroom guardrails; weak cells pass the minimum scheduled-service and V2V/active-float preservation guardrails but miss at least one stronger condition; failed cells miss one or more minimum guardrails. In strong and weak cells, cell text lists passed guardrails; in failed cells, it lists failed guardrails. Here svc means scheduled service, V2V means the voucher-to-voucher preservation guardrail, and head means issuer headroom. Cell labels summarize simulated guardrail status, not empirical field outcomes.}
\label{fig:edge-diagnostic-outcome-grid}
\end{figure}

\begin{figure}[htbp]
\centering
\includegraphics[width=0.9\textwidth]{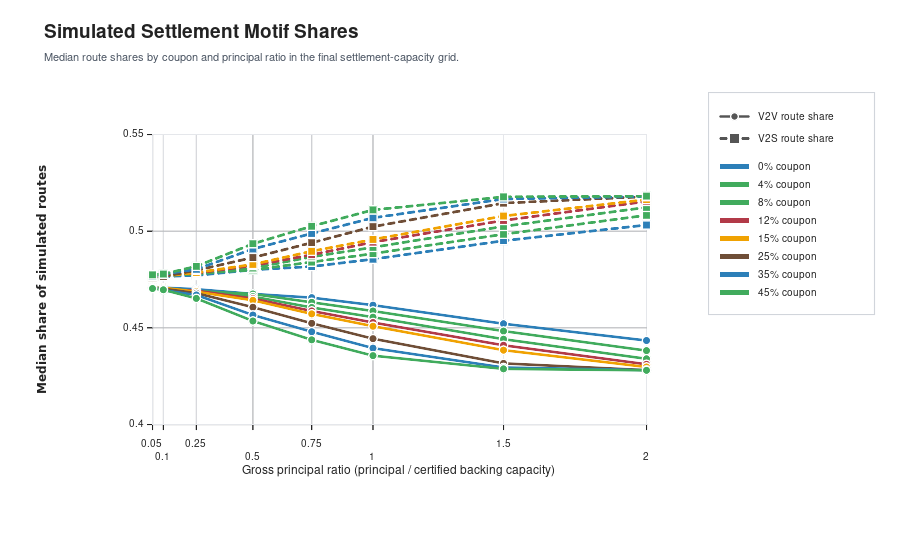}
\caption{Final settlement-capacity grid: simulated settlement motif shares across coupon and principal stress. The figure shows whether stable-seeking and voucher-to-voucher circulation move coherently as settlement pressure changes. It is a model-output diagnostic of settlement composition, not an empirical welfare, ecological, or legal-compliance outcome.}
\label{fig:edge-diagnostic-observed-motif-shift}
\end{figure}
\FloatBarrier

Figure~\ref{fig:edge-diagnostic-observed-motif-shift} clarifies why the frontier is not only a solvency or pass/fail grid. As principal stress rises, median simulated route composition shifts toward stable-seeking routes and away from voucher-to-voucher routes, even though the V2V value-preservation guardrail remains inside the material-decline tolerance across the tested grid. The diagnostic therefore supports reading the frontier as a settlement-attention and service-pressure test rather than as a recommendation to maximize principal or coupon.

The 1.25 headroom-multiple reference is a conservative diagnostic threshold rather than a legal covenant or estimated required reserve. Headroom is therefore interpreted as candidate operating and risk capacity, not as a proven sustainability reserve for training, monitoring, certification, audit, governance, legal costs, insurance claims, or first-loss protection.

\section{Discussion and Limitations}
The reported frontier gives a conditional result, not a deployment rule. Scheduled bondholder payment clears in 57 of 64 tested cells and becomes the first visible boundary at higher principal and coupon stress. Voucher-to-voucher value remains within the pre-specified material-decline guardrail in all cells, and it increases relative to matched no-bond baselines in 24 cells. The strong subset contains 52 cells that clear scheduled service, preserve voucher-to-voucher and active-float guardrails, and retain at least 1.25 median issuer headroom. This pattern supports the paper's mechanism claim under the reported assumptions: repayable liquidity can expand modeled producer-credit proxies and repayment pathways under productive-capacity assumptions while preserving local voucher circulation when issuer liability, pool limits, and service-resource rules remain separated.

The failures are equally informative. When cells fail, the first binding pressure is scheduled-service coverage rather than an immediate modeled collapse of voucher-to-voucher circulation. This matters because it locates the highest simulated stress at the formal issuer layer. It does not prove that local circulation would remain safe under every legal, governance, welfare, or cultural condition. It shows that, under the calibrated model, debt-service pressure can become binding before the protected local-circulation guardrail fails.

The result links two obligation layers without collapsing them. A bond advances capital now and obliges an issuer to repay over time under instrument terms. A voucher held by a pool may circulate under local redemption, fulfillment, and repair rules rather than securities law. Commitment pools create additional settlement paths: a voucher can be purchased, swapped, borrowed, routed, returned, redeemed, cleared, or repaired. The simulations provide conditional mechanism evidence that repayable external capital can support this multilateral obligation mobility when the issuer remains liable and local pool rules preserve curation, limits, exchange, and repair.

Grant-liquidity or philanthropic support is discussed here only as context, not as a separately simulated scenario. The empirical reports and transaction diagnostics provide an evidence infrastructure for monitoring local activity and settlement motifs, while causal welfare, productivity, ecological, dependency, and cultural-continuity claims require separate study. The frontier is read here as a test of settlement-capacity capture risk, not as a recommendation to maximize coupon rates or principal. In this framework, regenerative-bond finance is not simply more capital; it is capital within the carrying capacity of a settlement commons. This carrying capacity is treated as empirical and context-specific, not as a universal constant.

The limitations are substantive. Transaction volume by itself does not prove welfare, ecological regeneration, cultural legitimacy, or legal compliance. The empirical Sarafu material adds verified-report and audit evidence for reported activities, fulfillment records, pool activity, and calibration moments; those data support component-level evidence and monitoring design, not a standalone causal estimate of welfare or ecological impact. Token taxonomy, unit-of-account review, redemption mapping, field-report verification, and legal classification remain necessary before implementation claims can move beyond mechanism plausibility. Issuer sustainability also requires an explicit operating-cost and risk-capital layer, because headroom above scheduled debt service is only candidate capacity for training, monitoring, certification, audit, governance, and first-loss reserves.

Any deployment would need pre-registered ecological, welfare, governance, and settlement outcomes, including matched comparison pools or staggered rollout where feasible. The paper's contribution is not a claim that every proposed regenerative bond already works. It is a claim that such instruments can be specified, bounded, simulated, tested against failure conditions, and governed as a distinct settlement-architecture problem. Failure conditions include weaker repayment closure, lower route success, delayed redemption, insufficient eligible service resources, degraded voucher circulation, or local governance rejection. The evidence supports empirical calibration, within-sample calibration fit, validated report-linked component evidence, and a final 64-cell settlement-capacity frontier. Independent validation, jurisdiction-specific legal classification, ecological and welfare verification, and a bounded field test remain necessary before implementation claims can go beyond mechanism plausibility.

\section*{Data, Code Availability, and Positionality}
The public anonymized Sarafu Network dataset used for empirical calibration is archived on Zenodo \citep{ruddick2026sarafuDataset}. The version DOI \url{https://doi.org/10.5281/zenodo.20120190} identifies the archival dataset used for this paper; the concept DOI \url{https://doi.org/10.5281/zenodo.20120189} resolves to the dataset record family. The calibration and Monte Carlo simulation code is available in the public Cosmo Local Credit simulator repository \citep{cosmoLocalCreditSim}. The published source package should include an ancillary reproducibility note listing the simulator commit or release tag, runner command, configuration file, random-seed policy, and checksums for generated tables and figures. The paper uses derived and aggregate analysis outputs and does not expose personal contact fields, raw addresses, raw report text, tag text, or sensitive linkage fields.

Claims about reciprocal labor, rotating labor, savings-credit, and mutual-aid motifs are grounded in the cited comparative literature and in the Sarafu practitioner record, and are used here as analytical motifs rather than exhaustive local ethnographic classifications. The authors' practitioner involvement creates access to field experience and implementation details, but also creates risks of confirmation bias and overclaiming. The paper therefore separates practitioner design knowledge, verified-report evidence, empirical calibration, and Monte Carlo stress testing from broader welfare, legal, and ecological claims. The submission package provides privacy-safe aggregate outputs, checksums, and ancillary notes for the reported tables and figures; full independent reruns require the cited public simulator state, runner configuration, seed policy, and calibration inputs, not raw participant-level records.

\bibliographystyle{achicago}
\bibliography{regenerative_bonds}

\begin{thebibliography}{}

\bibitem[\UnexpandableProtect\SCcite{Acemoglu, Ozdaglar, and
  Tahbaz-Salehi}{2015}]{acemoglu2015systemic}
Acemoglu, Daron, Asuman Ozdaglar, and Alireza Tahbaz-Salehi.
\newblock 2015.
\newblock \Wrapquotes{Systemic Risk and Stability in Financial Networks}.
\newblock \textitswitch{American Economic Review}  105 (2): 564--608.
\newblock \url{https://doi.org/10.1257/aer.20130456}.

\bibitem[\UnexpandableProtect\SCcite{Anderson and
  Baland}{2002}]{anderson2002economics}
Anderson, Siwan, and Jean-Marie Baland.
\newblock 2002.
\newblock \Wrapquotes{The Economics of {ROSCAs} and Intrahousehold Resource
  Allocation}.
\newblock \textitswitch{The Quarterly Journal of Economics}  117 (3): 963--995.
\newblock \url{https://doi.org/10.1162/003355302760193931}.

\bibitem[\UnexpandableProtect\SCcite{Ardener}{1964}]{ardener1964comparative}
Ardener, Shirley.
\newblock 1964.
\newblock \Wrapquotes{The Comparative Study of Rotating Credit Associations}.
\newblock \textitswitch{The Journal of the Royal Anthropological Institute of
  Great Britain and Ireland}  94 (2): 201--229.
\newblock \url{https://doi.org/10.2307/2844382}.

\bibitem[\UnexpandableProtect\SCcite{Barde and van~der
  Hoog}{2017}]{barde2017abm}
Barde, Sylvain, and Sander van~der Hoog.
\newblock 2017.
\newblock \Wrapquotes{An Empirical Validation Protocol for Large-Scale
  Agent-Based Models}.
\newblock Discussion paper KDPE 1712, University of Kent School of Economics.
\newblock \url{https://ideas.repec.org/p/ukc/ukcedp/1712.html}. Accessed
  2026-05-11.

\bibitem[\UnexpandableProtect\SCcite{Besley, Coate, and
  Loury}{1993}]{besley1993economics}
Besley, Timothy, Stephen Coate, and Glenn Loury.
\newblock 1993.
\newblock \Wrapquotes{The Economics of Rotating Savings and Credit
  Associations}.
\newblock \textitswitch{The American Economic Review}  83 (4): 792--810.
\newblock \url{https://www.jstor.org/stable/2117579}. Accessed 2026-05-11.

\bibitem[\UnexpandableProtect\SCcite{Bouman}{1995}]{bouman1995roscas}
Bouman, F. J.~A.
\newblock 1995.
\newblock \Wrapquotes{Rotating and Accumulating Savings and Credit
  Associations: A Development Perspective}.
\newblock \textitswitch{World Development}  23 (3): 371--384.
\newblock \url{https://doi.org/10.1016/0305-750X(94)00141-K}.

\bibitem[\UnexpandableProtect\SCcite{Brunnermeier and
  Pedersen}{2009}]{brunnermeier2009}
Brunnermeier, Markus~K., and Lasse~Heje Pedersen.
\newblock 2009.
\newblock \Wrapquotes{Market Liquidity and Funding Liquidity}.
\newblock \textitswitch{Review of Financial Studies}  22 (6): 2201--2238.
\newblock \url{https://doi.org/10.1093/rfs/hhn098}.

\bibitem[\UnexpandableProtect\SCcite{Carroll et~al.}{2020}]{carroll2020care}
Carroll, Stephanie~Russo, Ibrahim Garba, Oscar~L. Figueroa-Rodriguez, Jarita
  Holbrook, Raymond Lovett, Simeon Materechera, Mark Parsons, Kay Raseroka,
  Desi Rodriguez-Lonebear, Robyn Rowe, Rodrigo Sara, Jennifer~D. Walker, Jane
  Anderson, and Maui Hudson.
\newblock 2020.
\newblock \Wrapquotes{The {CARE} Principles for Indigenous Data Governance}.
\newblock \textitswitch{Data Science Journal}  19:43.
\newblock \url{https://doi.org/10.5334/dsj-2020-043}.

\bibitem[\UnexpandableProtect\SCcite{{Committee on Payment and Settlement
  Systems} and {Technical Committee of IOSCO}}{2012}]{cpmiiosco2012}
{Committee on Payment and Settlement Systems}, and {Technical Committee of
  IOSCO}.
\newblock 2012.
\newblock \Wrapquotes{Principles for Financial Market Infrastructures}.
\newblock Technical Report, Bank for International Settlements, Basel.
\newblock \url{https://www.bis.org/cpmi/publ/d101.htm}. Accessed 2026-05-11.

\bibitem[\UnexpandableProtect\SCcite{{Cosmo Local
  Credit}}{2026}]{cosmoLocalCreditSim}
{Cosmo Local Credit}.
\newblock 2026.
\newblock sim: Agent-Based Simulation and Monte Carlo Calibration Code.
\newblock \url{https://github.com/cosmo-local-credit/sim}. Accessed 2026-05-23.

\bibitem[\UnexpandableProtect\SCcite{{Cosmo-Local
  Credit}}{2026}]{clcProtocolOverview}
{Cosmo-Local Credit}.
\newblock 2026.
\newblock Protocol Overview.
\newblock \url{https://cosmolocal.credit/protocol/overview/}. Accessed
  2026-05-11.

\bibitem[\UnexpandableProtect\SCcite{Duffie and Zhu}{2011}]{duffie2011}
Duffie, Darrell, and Haoxiang Zhu.
\newblock 2011.
\newblock \Wrapquotes{Does a Central Clearing Counterparty Reduce Counterparty
  Risk?}
\newblock \textitswitch{Review of Asset Pricing Studies}  1 (1): 74--95.
\newblock \url{https://doi.org/10.1093/rapstu/rar001}.

\bibitem[\UnexpandableProtect\SCcite{Fleischman, Dini, and
  Littera}{2020}]{fleischman2020liquidity}
Fleischman, Toma{\v z}, Paolo Dini, and Giuseppe Littera.
\newblock 2020.
\newblock \Wrapquotes{Liquidity-Saving through Obligation-Clearing and Mutual
  Credit: An Effective Monetary Innovation for {SMEs} in Times of Crisis}.
\newblock \textitswitch{Journal of Risk and Financial Management}  13 (12):
  295.
\newblock \url{https://doi.org/10.3390/jrfm13120295}.

\bibitem[\UnexpandableProtect\SCcite{Forastiere, Prete, and
  Sciabolazza}{2024}]{forastiere2024continuous}
Forastiere, Laura, Davide~Del Prete, and Valerio~Leone Sciabolazza.
\newblock 2024.
\newblock \Wrapquotes{Causal Inference on Networks under Continuous Treatment
  Interference}.
\newblock \textitswitch{Social Networks}  76:88--111.
\newblock \url{https://doi.org/10.1016/j.socnet.2023.07.005}.

\bibitem[\UnexpandableProtect\SCcite{Fullerton}{2015}]{fullerton2015regenerative}
Fullerton, John.
\newblock 2015.
\newblock \Wrapquotes{Regenerative Capitalism: How Universal Principles and
  Patterns Will Shape Our New Economy}.
\newblock Technical Report, Capital Institute.
\newblock \url{https://capitalinstitute.org/regenerative-capitalism/}. Accessed
  2026-05-11.

\bibitem[\UnexpandableProtect\SCcite{Geertz}{1962}]{geertz1962middle}
Geertz, Clifford.
\newblock 1962.
\newblock \Wrapquotes{The Rotating Credit Association: A {``}Middle Rung{''} in
  Development}.
\newblock \textitswitch{Economic Development and Cultural Change}  10 (3):
  241--263.
\newblock \url{https://doi.org/10.1086/449960}.

\bibitem[\UnexpandableProtect\SCcite{{GOV.UK}}{2026}]{govuk2026socialoutcomes}
{GOV.UK}.
\newblock 2026.
\newblock Social Outcomes Partnerships.
\newblock Updated 2026-03-05.
  \url{https://www.gov.uk/guidance/social-impact-bonds}. Accessed 2026-05-11.

\bibitem[\UnexpandableProtect\SCcite{Graeber}{2011}]{graeber2011}
Graeber, David.
\newblock 2011.
\newblock \textitswitch{Debt: The First 5,000 Years}.
\newblock Brooklyn, NY: Melville House.

\bibitem[\UnexpandableProtect\SCcite{{Grassroots
  Economics}}{2026a}]{cicStackDocs}
{Grassroots Economics}.
\newblock 2026a.
\newblock {CIC Stack Docs}.
\newblock \url{https://github.com/grassrootseconomics/cic-stack-docs}. Accessed
  2026-05-11.

\bibitem[\UnexpandableProtect\SCcite{{Grassroots
  Economics}}{2026b}]{erc20PoolRepo}
\SCduplicate{{Grassroots Economics}{}}.
\newblock 2026b.
\newblock erc20-pool.
\newblock \url{https://github.com/grassrootseconomics/erc20-pool}. Accessed
  2026-05-11.

\bibitem[\UnexpandableProtect\SCcite{Ingham}{2004}]{ingham2004}
Ingham, Geoffrey.
\newblock 2004.
\newblock \textitswitch{The Nature of Money}.
\newblock Cambridge: Polity Press.

\bibitem[\UnexpandableProtect\SCcite{{International Capital Market
  Association}}{2025a}]{icma2025gbp}
{International Capital Market Association}.
\newblock 2025a.
\newblock Green Bond Principles: Voluntary Process Guidelines for Issuing Green
  Bonds.
\newblock Updated June 2025.
  \url{https://www.icmagroup.org/sustainable-finance/the-principles-guidelines-and-handbooks/green-bond-principles-gbp}.
  Accessed 2026-05-11.

\bibitem[\UnexpandableProtect\SCcite{{International Capital Market
  Association}}{2025b}]{icma2025sbp}
\SCduplicate{{International Capital Market Association}{}}.
\newblock 2025b.
\newblock Social Bond Principles: Voluntary Process Guidelines for Issuing
  Social Bonds.
\newblock Updated June 2025.
  \url{https://www.icmagroup.org/sustainable-finance/the-principles-guidelines-and-handbooks/social-bond-principles-sbp/}.
  Accessed 2026-05-11.

\bibitem[\UnexpandableProtect\SCcite{{International Monetary Fund} and {World
  Bank}}{2014}]{worldbankimf2014debt}
{International Monetary Fund}, and {World Bank}.
\newblock 2014.
\newblock \Wrapquotes{Revised Guidelines for Public Debt Management}.
\newblock Technical Report, International Monetary Fund and World Bank.
\newblock
  \url{https://www.imf.org/en/publications/policy-papers/issues/2016/12/31/revised-guidelines-for-public-debt-management-pp4855}.
  Accessed 2026-05-11.

\bibitem[\UnexpandableProtect\SCcite{Kropotkin}{1902}]{kropotkin1902}
Kropotkin, Peter.
\newblock 1902.
\newblock \textitswitch{Mutual Aid: A Factor of Evolution}.
\newblock London: William Heinemann.

\bibitem[\UnexpandableProtect\SCcite{Leung}{2022}]{leung2022interference}
Leung, Michael~P.
\newblock 2022.
\newblock \Wrapquotes{Causal Inference Under Approximate Neighborhood
  Interference}.
\newblock \textitswitch{Econometrica}  90 (1): 267--293.
\newblock \url{https://doi.org/10.3982/ECTA17841}.

\bibitem[\UnexpandableProtect\SCcite{Mauss}{2002}]{mauss2002}
Mauss, Marcel.
\newblock 2002.
\newblock \textitswitch{The Gift: The Form and Reason for Exchange in Archaic
  Societies}.
\newblock London: Routledge.

\bibitem[\UnexpandableProtect\SCcite{Mehrling}{2011}]{mehrling2011}
Mehrling, Perry.
\newblock 2011.
\newblock \textitswitch{The New Lombard Street: How the Fed Became the Dealer
  of Last Resort}.
\newblock Princeton, NJ: Princeton University Press.

\bibitem[\UnexpandableProtect\SCcite{Messerschmidt}{1981}]{messerschmidt1981nogar}
Messerschmidt, Donald~A.
\newblock 1981.
\newblock \Wrapquotes{Nogar and Other Traditional Forms of Cooperation in
  Nepal: Significance for Development}.
\newblock \textitswitch{Human Organization}  40 (1): 40--47.
\newblock \url{https://doi.org/10.17730/humo.40.1.k7811l2478272681}.

\bibitem[\UnexpandableProtect\SCcite{Minsky}{1986}]{minsky1986}
Minsky, Hyman~P.
\newblock 1986.
\newblock \textitswitch{Stabilizing an Unstable Economy}.
\newblock New Haven, CT: Yale University Press.

\bibitem[\UnexpandableProtect\SCcite{{OECD}}{2025}]{oecd2025globaldebt}
{OECD}.
\newblock 2025.
\newblock \Wrapquotes{Global Debt Report 2025: Financing Growth in a
  Challenging Debt Market Environment}.
\newblock Technical Report, OECD Publishing, Paris.
\newblock \url{https://doi.org/10.1787/8ee42b13-en}. Accessed 2026-05-11.

\bibitem[\UnexpandableProtect\SCcite{Ostrom}{1990}]{ostrom1990}
Ostrom, Elinor.
\newblock 1990.
\newblock \textitswitch{Governing the Commons: The Evolution of Institutions
  for Collective Action}.
\newblock Cambridge: Cambridge University Press.

\bibitem[\UnexpandableProtect\SCcite{Polanyi}{1944}]{polanyi1944}
Polanyi, Karl.
\newblock 1944.
\newblock \textitswitch{The Great Transformation: The Political and Economic
  Origins of Our Time}.
\newblock New York, NY: Farrar \& Rinehart.

\bibitem[\UnexpandableProtect\SCcite{Ruddick and
  Sohail~Azim}{2026}]{ruddick2026sarafuDataset}
Ruddick, William, and Mohamed Sohail~Azim.
\newblock 2026.
\newblock Sarafu Network Public Dataset.
\newblock Dataset. Zenodo. Version DOI:
  \url{https://doi.org/10.5281/zenodo.20120190}. Concept DOI:
  \url{https://doi.org/10.5281/zenodo.20120189}. CC BY 4.0. Accessed
  2026-05-17.

\bibitem[\UnexpandableProtect\SCcite{Ruddick}{2023}]{ruddick2023letters}
Ruddick, William~O.
\newblock 2023.
\newblock \Wrapquotes{Letters from the Field: Commitment Pooling--An Economic
  Protocol Inspired by Ancestral Wisdom}.
\newblock \textitswitch{International Journal of Community Currency Research}
  27:54--79.
\newblock \url{https://doi.org/10.15133/j.ijccr.2023.004}.

\bibitem[\UnexpandableProtect\SCcite{Ruddick}{2026a}]{ruddick2026routegraphs}
\SCduplicate{Ruddick, William~O.{}}.
\newblock 2026a, April.
\newblock Commitment-Pool Route Graphs for Finance and Mutual Aid.
\newblock Available at SSRN: \url{https://ssrn.com/abstract=6606438} or
  \url{https://doi.org/10.2139/ssrn.6606438}. Accessed 2026-05-17.

\bibitem[\UnexpandableProtect\SCcite{Ruddick}{2026b}]{ruddick2026proto}
\SCduplicate{Ruddick, William~O.{}}.
\newblock 2026b.
\newblock \Wrapquotes{Proto-Social Infrastructure and Stewardship of Commitment
  Pooling}.
\newblock \textitswitch{International Journal of Community Currency Research}
  30 (1): 80--98.
\newblock \url{https://www.ijccr.net/article/view/9512}. Accessed 2026-05-11.

\bibitem[\UnexpandableProtect\SCcite{Sahlins}{1972}]{sahlins1972}
Sahlins, Marshall.
\newblock 1972.
\newblock \textitswitch{Stone Age Economics}.
\newblock Chicago: Aldine-Atherton.

\bibitem[\UnexpandableProtect\SCcite{Smith}{2012}]{smith2012decolonizing}
Smith, Linda~Tuhiwai.
\newblock 2012.
\newblock \textitswitch{Decolonizing Methodologies: Research and Indigenous
  Peoples}.
\newblock 2.
\newblock London: Zed Books.

\bibitem[\UnexpandableProtect\SCcite{Wang}{2019}]{wang2019rola}
Wang, Shun.
\newblock 2019.
\newblock \Wrapquotes{Social Capital and Rotating Labor Associations in Rural
  China}.
\newblock \textitswitch{China Economic Review}  53:243--253.
\newblock \url{https://doi.org/10.1016/j.chieco.2018.09.013}.

\end{thebibliography}


\begin{thebibliography}{}

\bibitem[\UnexpandableProtect\SCcite{Andersen et~al.}{2001}]{andersen2001ron}
Andersen, David~G., Hari Balakrishnan, M.~Frans Kaashoek, and Robert Morris.
\newblock 2001.
\newblock \Wrapquotes{Resilient Overlay Networks}.
\newblock \textitswitch{Proceedings of the 18th {ACM} Symposium on Operating
  Systems Principles}.
\newblock 131--145.
\newblock \url{https://doi.org/10.1145/502034.502048}.

\bibitem[\UnexpandableProtect\SCcite{{Committee on Payment and Settlement
  Systems} and {Technical Committee of IOSCO}}{2012}]{cpmiiosco2012}
{Committee on Payment and Settlement Systems}, and {Technical Committee of
  IOSCO}.
\newblock 2012.
\newblock \Wrapquotes{Principles for Financial Market Infrastructures}.
\newblock Technical Report, Bank for International Settlements, Basel.
\newblock \url{https://www.bis.org/cpmi/publ/d101.htm}. Accessed 2026-05-11.

\bibitem[\UnexpandableProtect\SCcite{{Cosmo Local
  Credit}}{2026a}]{cosmoLocalCreditDocs}
{Cosmo Local Credit}.
\newblock 2026a.
\newblock docs: Cosmo-Local Credit Documentation.
\newblock \url{https://github.com/cosmo-local-credit/docs}. Accessed
  2026-05-23.

\bibitem[\UnexpandableProtect\SCcite{{Cosmo Local
  Credit}}{2026b}]{cosmoLocalCreditProtocol}
\SCduplicate{{Cosmo Local Credit}{}}.
\newblock 2026b.
\newblock protocol: Cosmo-Local Credit Protocol Repository.
\newblock \url{https://github.com/cosmo-local-credit/protocol}. Accessed
  2026-05-23.

\bibitem[\UnexpandableProtect\SCcite{{Cosmo Local
  Credit}}{2026c}]{cosmoLocalCreditSim}
\SCduplicate{{Cosmo Local Credit}{}}.
\newblock 2026c.
\newblock sim: Agent-Based Simulation and Monte Carlo Calibration Code.
\newblock \url{https://github.com/cosmo-local-credit/sim}. Accessed 2026-05-23.

\bibitem[\UnexpandableProtect\SCcite{Eisenberg and
  Noe}{2001}]{eisenberg2001systemic}
Eisenberg, Larry, and Thomas~H. Noe.
\newblock 2001.
\newblock \Wrapquotes{Systemic Risk in Financial Systems}.
\newblock \textitswitch{Management Science}  47 (2): 236--249.
\newblock \url{https://doi.org/10.1287/mnsc.47.2.236.9835}.

\bibitem[\UnexpandableProtect\SCcite{Eppstein}{1998}]{eppstein1998kshortest}
Eppstein, David.
\newblock 1998.
\newblock \Wrapquotes{Finding the {$k$} Shortest Paths}.
\newblock \textitswitch{{SIAM} Journal on Computing}  28 (2): 652--673.
\newblock \url{https://doi.org/10.1137/S0097539795290477}.

\bibitem[\UnexpandableProtect\SCcite{Fleischman, Dini, and
  Littera}{2020}]{fleischman2020liquidity}
Fleischman, Toma{\v z}, Paolo Dini, and Giuseppe Littera.
\newblock 2020.
\newblock \Wrapquotes{Liquidity-Saving through Obligation-Clearing and Mutual
  Credit: An Effective Monetary Innovation for {SMEs} in Times of Crisis}.
\newblock \textitswitch{Journal of Risk and Financial Management}  13 (12):
  295.
\newblock \url{https://doi.org/10.3390/jrfm13120295}.

\bibitem[\UnexpandableProtect\SCcite{{Grassroots
  Economics}}{2026}]{sarafuNetworkRepo}
{Grassroots Economics}.
\newblock 2026.
\newblock Sarafu Network Repository.
\newblock \url{https://github.com/grassrootseconomics/sarafu.network}. Accessed
  2026-05-23.

\bibitem[\UnexpandableProtect\SCcite{Martins}{1984}]{martins1984multicriteria}
Martins, Ernesto Q.~V.
\newblock 1984.
\newblock \Wrapquotes{On a Multicriteria Shortest Path Problem}.
\newblock \textitswitch{European Journal of Operational Research}  16 (2):
  236--245.
\newblock \url{https://doi.org/10.1016/0377-2217(84)90077-8}.

\bibitem[\UnexpandableProtect\SCcite{Ruddick and
  Sohail~Azim}{2026}]{ruddick2026sarafuDataset}
Ruddick, William, and Mohamed Sohail~Azim.
\newblock 2026.
\newblock Sarafu Network Public Dataset.
\newblock Dataset. Zenodo. Version DOI:
  \url{https://doi.org/10.5281/zenodo.20120190}. Concept DOI:
  \url{https://doi.org/10.5281/zenodo.20120189}. CC BY 4.0. Accessed
  2026-05-17.

\bibitem[\UnexpandableProtect\SCcite{Tassiulas and
  Ephremides}{1992}]{tassiulas1992stability}
Tassiulas, Leandros, and Anthony Ephremides.
\newblock 1992.
\newblock \Wrapquotes{Stability Properties of Constrained Queueing Systems and
  Scheduling Policies for Maximum Throughput in Multihop Radio Networks}.
\newblock \textitswitch{{IEEE} Transactions on Automatic Control}  37 (12):
  1936--1948.
\newblock \url{https://doi.org/10.1109/9.182479}.

\bibitem[\UnexpandableProtect\SCcite{Yen}{1971}]{yen1971kshortest}
Yen, Jin~Y.
\newblock 1971.
\newblock \Wrapquotes{Finding the {$K$} Shortest Loopless Paths in a Network}.
\newblock \textitswitch{Management Science}  17 (11): 712--716.
\newblock \url{https://doi.org/10.1287/mnsc.17.11.712}.

\end{thebibliography}

\end{document}


\maketitle

\section{Purpose of the Supplement}
This supplement preserves the technical and evidentiary details that support the main article. It includes field-history notes, design invariants, model-object mappings, measurement bridges, detailed simulation assumptions, calibration-fit tables, background template counterfactuals, glossary terms, and technical source integration. These materials are not an independent proof that regenerative bonds work. They specify how the mechanism is represented, bounded, calibrated, and stress tested.

\section{Field-History Notes}
The main article treats ROLA-like and ROSCA-like as analytical motifs, not as direct cultural classifications. The terminology is grounded in the comparative ROLA, ROSCA, mutual-aid, gift, commons, and institutional literatures cited in the main text and in the Grassroots Economics practitioner record. Field accounts from Grassroots Economics describe communities in which savings-credit and mutual-aid practices persisted while some reciprocal labor, local exchange, and rotating-support practices had weakened, changed form, or become less institutionally visible. Community asset voucher training, resource mapping, membership agreements, mutual redemption exercises, and initial program liquidity then helped participants express redeemable commitments in pool form. The paper uses these accounts as field-history context while grounding its quantitative claims in observed pool activity, voucher circulation, stable-credit motifs, verified-report evidence, calibration fit, and counterfactual frontier tests.

Sarafu Network is not treated as a historical regenerative bond. Philanthropic or programmatic support is the observed catalyst in the empirical record; repayable bond liquidity is the counterfactual catalyst in the simulation. The reported model does not begin from a voucher-inactive ROSCA-like network or test whether bond-supported credit can create the voucher-capable layer itself. It studies a voucher-capable credit-pool substrate and asks whether repayable liquidity preserves or amplifies ROLA-like circulation while maintaining issuer responsibility for formal debt.

\section{Design Invariants}
\begingroup
\small
\setlength{\tabcolsep}{3pt}
\begin{longtable}{P{0.09\textwidth}P{0.40\textwidth}P{0.25\textwidth}P{0.16\textwidth}}
\caption{Full design invariants and test paths}
\label{tab:supp-design-invariants}\\
\toprule
ID & Statement & Protection & Test path \\
\midrule
\endfirsthead
\toprule
ID & Statement & Protection & Test path \\
\midrule
\endhead
L1 & Formal bondholder claims and local relational obligations remain distinct. & Prevents hidden household guarantee. & Legal structure review and simulator invariant. \\
L2 & Local stewards retain curation authority over admissible commitments. & Prevents cultural flattening and central capture. & Governance audit and override stress. \\
L3 & Advances remain bounded by valuation, evidence, caps, and haircuts. & Prevents over-issuance and collateral fantasy. & Limit and haircut stress tests. \\
L4 & Repair paths are explicit: where substitution, routing, deferment, guarantee, insurance, re-entry, or loss recognition are used, each has a trigger, authority, evidence rule, state update, and exit or re-entry condition. & Preserves multilateral repair beyond bilateral cash repayment. & Cash-only versus repair-enabled ablation. \\
L5 & Routing can improve settlement only when admissibility, inventory, limits, fees, and guards pass. & Avoids magical interoperability claims. & Route graph and failed-route analysis. \\
L6 & Operational solvency depends on cash-usable fee revenue, not gross in-kind fee volume. & Avoids false service-capacity claims. & Cash-eligible share sensitivity. \\
L7 & Issuance, transfer, settlement, fee allocation, and repair generate receipts or evidence records. & Enables verification and dispute handling. & Software invariant and reporting audit. \\
L8 & Debt service cannot depend primarily on scarce cash extraction from vulnerable households. & Protects the non-extraction boundary. & Household cash-stress and coercion flags. \\
L9 & Pool-level obligation mobility is distinct from formal bondholder collateral. & Preserves tradable local settlement without hidden guarantees. & Exposure-shift and service-cash accounting audit. \\
\bottomrule
\end{longtable}
\endgroup

\section{Publication Scope and Jurisdictional Due Diligence}
The paper is finalized as an academic mechanism, evidence, and simulation study. It does not offer securities, tax, accounting, investment, or implementation advice. Jurisdiction-specific due diligence is a deployment requirement for any real issuer, instrument, reserve, settlement asset, fee pathway, or voucher system. Table \ref{tab:supp-due-diligence-matrix} consolidates the legal, accounting, tax, and governance checks that would be required before implementation.

\begingroup
\small
\setlength{\tabcolsep}{3pt}
\begin{longtable}{P{0.25\textwidth}P{0.33\textwidth}P{0.32\textwidth}}
\caption{Jurisdictional due-diligence matrix for deployment}
\label{tab:supp-due-diligence-matrix}\\
\toprule
Domain & Paper treatment & Deployment due diligence \\
\midrule
\endfirsthead
\toprule
Domain & Paper treatment & Deployment due diligence \\
\midrule
\endhead
Redeemable commitments and vouchers & Modeled and empirically observed as governed promises or voucher representations, not assumed to be securities-law debt. & Classify each instrument under contract, voucher, e-money, payments, consumer-protection, cooperative, tax, and securities rules. \\
Stable-value settlement assets & Treated as data and simulator settlement categories, not as legal tender, deposits, insured claims, or accounting cash equivalents. & Confirm legal status, reserve backing, custody, convertibility, accounting recognition, tax treatment, and payment-system compliance. \\
Formal issuer authority & The framework requires a legally authorized issuer and keeps pools and households outside the bondholder claim. & Confirm debt-issuance authority, disclosure duties, authorization process, contingent-liability treatment, and non-guarantee boundaries for pools and participants. \\
Eligible service resources & Only legally available cash, stable value, reserves, recovered stable, or contractually available revenues can service formal debt. & Validate fee allocation, conversion, lockbox, waterfall, reserve-draw, tax, procurement, and public-finance rules. \\
Reserve and headroom accounting & $R^{pay}_t$ and $R^{rel}_t$ separate scheduled-service availability from unrestricted issuer headroom. & Specify reserve covenants, release tests, permitted draws, accounting treatment, audit evidence, and investor disclosure. \\
Shortfall and default & The model reports scheduled-service shortfalls before instrument-specific cure or restructuring terms. & Draft event-of-default, cure, grace, deferral, capitalization, acceleration, restructuring, and remedy provisions. \\
Simulation terms & Principal ratios, coupon targets, amortization, and lockbox assumptions are stress-test settings. & Convert any proposed transaction into actual term sheets, covenants, disclosures, risk factors, local approvals, and regulated offering documents where required. \\
\bottomrule
\end{longtable}
\endgroup

\section{Model Objects and Measurement Bridge}
\begingroup
\small
\setlength{\tabcolsep}{2pt}
\begin{longtable}{P{0.15\textwidth}P{0.22\textwidth}P{0.21\textwidth}P{0.19\textwidth}P{0.15\textwidth}}
\caption{Model objects, Sarafu implementation evidence, and Monte Carlo parameters}
\label{tab:supp-model-sarafu-monte-carlo}\\
\toprule
Model object & Sarafu implementation & Observable metric & Epistemic status & Monte Carlo parameter \\
\midrule
\endfirsthead
\toprule
Model object & Sarafu implementation & Observable metric & Epistemic status & Monte Carlo parameter \\
\midrule
\endhead
Voucher
& Voucher records, issuers, transfers, mints, burns, redemptions.
& Outstanding voucher stock, commitment-issuer-return proxy, redemption lag.
& Obligation proxy, not fulfillment proof.
& Supply, redemption probability, settlement latency. \\
\addlinespace
Pool
& Pool deposits, swaps, allowed-token records, token limits, exchange rates.
& Pool cadence, credit-to-backing ratio, deposit-to-limit ratio, active voucher count.
& Implemented curation, valuation, limitation, and exchange traces.
& Pool caps, value index, swap attempts, liquidity inventory. \\
\addlinespace
ROLA-like circulation
& Voucher-to-voucher swaps and own-voucher-in-for-other-voucher-out motifs.
& ROLA-like pool-weeks, turn-taking recurrence, round-completion proxy.
& Behavioral motif, not cultural proof.
& Reciprocal exchange cadence, round completion, repair delay. \\
\addlinespace
ROSCA-like stable credit
& Stable-to-voucher and voucher-to-stable swaps, pool cash inflows, credit-limit snapshots.
& Borrow proxy, debt-removal proxy, stable share of flows.
& Pooled-credit motif unless rotation is directly established.
& Stable injection, borrowing rights, repayment rates. \\
\addlinespace
Liquidity injection
& Standalone backing deposits into pools and later swaps of injected assets.
& Injected stable/redeemable value, borrowed-stable proxy, unresolved stock.
& Diagnostic liquidity evidence.
& Bond proceeds allocation, liquidity mandate, leakage rate. \\
\addlinespace
Cosmo-local credit
& Shared voucher listings and pool-token graph components.
& Connected components, bridge vouchers, shared-voucher pool-pool edges.
& Settlement-path possibility, not route execution proof.
& Route graph density, routing pass rate, failed-route reasons. \\
\addlinespace
Impact evidence
& Voucher-linked field reports with status, verifier IDs, tags, and periods.
& Approved reports, verified reports, report-period activity alignment.
& Outcome-evidence layer requiring local review.
& Report arrival, verification rate, ecological/social task completion. \\
\addlinespace
Formal bond layer
& Not historically implemented in Sarafu.
& Simulated issuer principal, eligible local commitment-pool deployment, producer debt, recovered stable, scheduled service, and explicit clearing.
& Counterfactual simulation layer.
& Bond principal, coupon, scheduled service, pool recovery, cash headroom, and binding guardrails. \\
\bottomrule
\end{longtable}
\endgroup

\begingroup
\small
\setlength{\tabcolsep}{2pt}
\renewcommand{\arraystretch}{1.08}
\begin{longtable}{P{0.16\textwidth}P{0.21\textwidth}P{0.19\textwidth}P{0.18\textwidth}P{0.18\textwidth}}
\caption{Measurement bridge from social bonds to formal regenerative-bond hypotheses}
\label{tab:social-financial-bond-measurement-bridge}\\
\toprule
Concept & Observable in Sarafu data & Empirical metric & Simulator parameter & Bond hypothesis \\
\midrule
\endfirsthead
\toprule
Concept & Observable in Sarafu data & Empirical metric & Simulator parameter & Bond hypothesis \\
\midrule
\endhead
Social bond / obligation
& Voucher issuance, pool swaps, issuer returns, burns
& Circulation volume, issuer-return proxy, redemption lag, unresolved voucher stock
& Commitment supply, redemption probability, settlement latency
& Local obligations can be measured without making households formal bond obligors. \\
\addlinespace
Commitment pool
& Voucher listings, limits, exchange rates, deposits, swaps, fees
& Credit-to-backing ratio, pool cadence, registry coverage, pool-week activity
& Pool caps, value index, fee rate, swap attempts
& Pool rules form the institutional membrane between social obligation and formal finance. \\
\addlinespace
ROLA-like mutual aid
& Repeated voucher-to-voucher turns among pool members, followed by issuer-return proxies
& ROLA-like pool-weeks, round-completion proxy, turn-taking balance, return lag
& Reciprocal-exchange cadence, fulfillment rate, repair delay
& Mutual-aid circulation can become empirically legible without being securitized. \\
\addlinespace
ROSCA-like stable credit
& Stable/cash deposits, stable-to-voucher swaps, voucher-to-stable swaps, credit limits
& Stable share, borrow proxy, debt-removal proxy, deposit-to-limit ratio
& Stable injection, borrowing rights, loan and repayment rates
& Stable liquidity can create borrowing rights and help retire obligations when redemption demand exists. \\
\addlinespace
Cosmo-local credit
& Shared vouchers across pools, pool-token components, multi-pool transactions
& Bridge vouchers, connected components, route-like swaps, concentration
& Route graph density, routing pass rate, clearing frequency
& Federation can increase settlement paths, but connectivity must also be stress-tested for fragility. \\
\addlinespace
Ecological and welfare evidence
& Voucher-linked field reports, report tags, report periods, verification status
& Approved and verified reports, report-activity alignment, tag-category counts
& Outcome-report arrival, verification rate, reporting lag
& Regenerative claims require a separate verified reporting layer beyond transaction traces. \\
\addlinespace
Formal financial bond
& Modeled eligible local commitment-pool deployment and recovery, not a direct Sarafu historical fact
& Recovered stable, service-cash headroom, settlement velocity, cash eligibility
& Gross bond principal, eligible local commitment pools, coupon, scheduled service
& Bond finance supports settlement capacity only if issuer repayment is explicit and not shifted to vulnerable participants. \\
\bottomrule
\end{longtable}
\endgroup

\FloatBarrier

\section{Detailed Sarafu Motifs}
ROLA-like circulation is measured as movable reciprocal commitment exchange inside pools. The main observed form is recurring voucher-to-voucher exchange, especially where participants appear to place their own voucher into a pool and draw another voucher out, followed by later commitment-issuer-return or burn evidence. The Kenya KES community-pool cohort reports 5,103 voucher-to-voucher pool swaps with 9,030.81 input value, 105 own-voucher-in-for-other-voucher-out proxy events with 1,040.28 value, and 7 ROLA-like pool-weeks across 7 pools.

The 7 ROLA-like pool-week count is a stricter recurrence or round-completion proxy than the raw voucher-to-voucher swap count. It therefore indicates that only a small subset of observed V2V activity satisfies the stricter motif definition; it is not the total amount of reciprocal commitment exchange in the cohort.

ROSCA-like stable-credit behavior appears when stable or cash-like assets enter pools and participants acquire borrowing or acceptance capacity through voucher deposits, pool limits, or stable withdrawals. The cohort reports 441 stable-to-voucher purchase events that act as debt-reduction proxies, with 10,376.00 KES-denominated value of cash or stable input. It also reports 2,148 voucher-to-stable borrowing or advance proxy events, with 39,656.26 KES-denominated value of cash or stable output. Mature voucher-to-stable borrow-proxy events later show commitment-issuer-return or burn support in 878 of 1,554 cases, a mature event support rate of 56.5 percent and a mature value support rate of 64.9 percent. Here mature means that the event has passed the pre-specified observation window used by the audit, so later commitment-issuer return or burn support could in principle be observed.

Stable liquidity is useful because it can settle obligations, remove voucher debt, and make pool credit usable. It is also a dependency risk. The stable-dependency calibration reports a 44.7 percent aggregate stable/cash share of observed KES cohort pool deposit, swap-in, and swap-out value under metadata-weighted voucher valuation, and a 60.8 percent share under the simulator-comparable strict convention that one individual voucher token equals one KES. Positive-net flow-balance stable share is much lower: 0.013 percent under metadata weighting and 0.027 percent under the strict one-voucher-equals-one-KES convention. These values are dependency diagnostics rather than welfare or legal-convertibility claims.

\subsection{Same-token return and producer-voucher integrity audits}
The same-token return audit is a pool-token accounting diagnostic. It treats each swap out of a pool as an open pool-token exposure and matches later swaps of the same token back into the same pool on a first-in-first-out basis, capped at the original outflow. The package includes the aggregate audit report and by-asset calibration CSV; both contain privacy-safe summaries rather than participant-level histories.

\begin{table}[htbp]
\centering
\caption{Same-pool same-token FIFO return audit}
\label{tab:supp-same-token-return-audit}
\small
\setlength{\tabcolsep}{4pt}
\begin{tabularx}{\textwidth}{P{0.30\textwidth}YYY}
\toprule
Asset class & Pool-token pairs & Swap-out value & Matched later in / coverage \\
\midrule
Cash/stable & 91 & 57,695.44 & 13,742.77 / 23.8\% \\
Redeemable voucher & 221 & 19,131.72 & 16,874.83 / 88.2\% \\
Internal voucher & 602 & 12,383.63 & 7,987.92 / 64.5\% \\
Overall & 914 & 89,210.79 & 38,605.51 / 43.3\% \\
\bottomrule
\end{tabularx}
\end{table}

The producer-voucher integrity audit is a separate boundary check, not a legal debt ledger. Its aggregate output reports 1,279 mint events; 99.7 percent of strict-unit mint value initially reaches producer wallets; producer-voucher pool-deposit logs total 20,272.10 KES-denominated value; observed producer-voucher credit limits total 67,049.58; producer-voucher swap-in value totals 44,649.35; and the pool-held voucher flow-balance proxy is 29,526.99. These quantities support the interpretation that the retained cohort provides pool-level exposure, limit, and settlement evidence while leaving individual legal obligations, fulfillment quality, and welfare outside the transaction audit.

\begin{table}[htbp]
\centering
\caption{Trailing-52-week actor-split stable-to-voucher calibration. Producer vouchers are Sarafu Network voucher representations of producer-issued redeemable commitments. Stable-to-voucher input means stable or cash-like value used to acquire or remove a voucher. Borrower self stable-to-own-voucher input is direct self-repayment proxy value. Other-producer input is producer stable used to acquire another producer's voucher. External non-producer input is stable-side demand from wallets outside the producer set. Producer stable reuse share compares producer stable-to-voucher reuse with producer voucher-to-stable receipts.}
\label{tab:supp-stable-actor-split-calibration}
\small
\setlength{\tabcolsep}{4pt}
\begin{tabularx}{\textwidth}{P{0.34\textwidth}YP{0.42\textwidth}}
\toprule
Calibration quantity & Value & Interpretation \\
\midrule
Borrower self stable-to-own-voucher input & 10.65 KES value/year & Direct observed self-repayment proxy over the calibration window. \\
Other-producer stable-to-voucher input & 415.747780 KES value/year & Empirically small producer-stable reuse into other producer vouchers. \\
External non-producer stable-to-voucher input & 1,790.510041 KES value/year & External buyer stable input; not automatically interpreted as household consumption. \\
Producer stable reuse share & 0.03248 & Producer stable-to-voucher reuse relative to producer voucher-to-stable receipts. \\
\bottomrule
\end{tabularx}
\end{table}

\section{Simulation Design Details}
\subsection{Agent and state representation}
The simulator contains traversable liquidity and settlement pools, a simulator class for rule-governed liquidity and settlement venues rather than a legal lending classification, plus private producer wallets, private consumer wallets, local steward groups, public agencies, a regenerative settlement facility, a formal issuer, and optional insurance or guarantor state. The reported topology uses 73 traversable liquidity and settlement pools, 996 producer wallets, 1,247 accepted-voucher member slots, and 462 external non-producer consumer wallets at calibrated-network scale. Producer vouchers are Sarafu Network representations of producer-issued redeemable commitments, as defined in the main paper. They can be accepted in multiple eligible pools according to the empirical overlap calibration. Producer and consumer wallets can start or receive a route, but they are not traversable swap or clearing venues for other agents. Agents issue commitments, request advances, fulfill, redeem, repay, repair, buy commitments, route swaps, receive repayment, adjust limits, pause listings, and record reports. The model separates transaction settlement from outcome verification.

\begingroup
\small
\setlength{\tabcolsep}{3pt}
\begin{longtable}{P{0.18\textwidth}P{0.37\textwidth}P{0.32\textwidth}}
\caption{Monte Carlo agents and actions}
\label{tab:supp-monte-carlo-agents}\\
\toprule
Class & State variables & Actions \\
\midrule
\endfirsthead
\toprule
Class & State variables & Actions \\
\midrule
\endhead
Household or producer
& Commitment supply, outstanding obligations, cash balance, fulfillment capacity, reputation, repair status.
& Issue commitment, request advance, fulfill, redeem, repay, repair. \\
Consumer or buyer
& Cash or settlement balance, needs, trust preferences, redemption demand.
& Buy commitment, route swap, redeem, report fulfillment. \\
Liquidity or settlement pool
& Liquidity, accepted commitment exposure, limits, fees, risk appetite.
& Advance accepted assets, accept commitments, route, receive repayment or closure. \\
Local steward group
& Curation rules, value references, limits, pause authority, repair decisions.
& List, delist, value, relimit, pause, adjudicate, record. \\
Formal issuer
& Bond principal, coupon scenario, maturity, eligible recovered stable, service headroom, disclosure duties.
& Issue bond, pay coupon, refinance, report. \\
\bottomrule
\end{longtable}
\endgroup

\subsection{Credit, routing, and redemption assumptions}
Producer credit is represented through commitment-pool accounting rather than through a conventional bank-loan object. When a producer swaps its own voucher for an accepted asset such as stable value or another voucher, the model treats the received asset as potential productive or settlement capacity and the receiving pool holds the producer's obligation. The empirical data provide component evidence for pool value, stable-credit interaction, report-linked activity, and repayment or closure proxies used in calibration. The frontier mechanism then treats both own-voucher-to-stable and own-voucher-to-voucher borrowing as modeled producer-credit channels with productive-capacity assumptions; it does not by itself prove that every advance increases real output, income, welfare, or ecological capacity. Modeled stable receipts arrive after a lag and can fund monthly stable-to-own-voucher self-repayment. Repayment or closure occurs when that producer voucher is removed from the creditor pool: the producer may route stable back to buy its own voucher, or another participant may acquire the voucher and accept the claim on the commitment issuer. If a producer reaches its pool-side own-voucher exposure limit, further own-voucher borrowing is suppressed rather than converted into artificial route failure. The reported results keep the static 5x deposit-based borrowing cap as an internal guardrail and audit diagnostic; borrowing-cap edges are sensitivity diagnostics rather than headline results. The broader frontier is therefore framed as a settlement-capacity risk: guardrail-pass cells expand modeled producer-credit proxies under productive-capacity assumptions without degrading ROLA-like circulation or issuer headroom, while failure cells show scheduled service, stable dependency, leakage, concentration, or settlement-attention pressure becoming binding.

The simulator includes a network-aware routing and clearing module. The detailed implementation is a network-aware scoring layer over bounded overlay or beam search and live pool validation. It constructs bounded candidate sets, searches with beam-like route expansion, scores edges using multiple objectives, validates execution against live pool limits and inventories, and optionally clears feasible cycles. In the reported topology, the executable graph traverses eligible pools only; producer and consumer private wallets are route endpoints rather than intermediate venues. This heuristic is in the same broad family as $k$-shortest-path candidate generation, multiobjective shortest-path routing, resilient overlay routing, adaptive or backpressure-style network control, and financial obligation-clearing or netting systems \citep{yen1971kshortest,eppstein1998kshortest,martins1984multicriteria,andersen2001ron,tassiulas1992stability,eisenberg2001systemic,cpmiiosco2012,fleischman2020liquidity}. The relevant falsification test is whether the frontier shrinks when route visibility, cache, beam width, affinity, or route-success thresholds are weakened.

Automatic or rapid third-party voucher redemption is a settlement-timing assumption. It reflects face-to-face exchange and the empirical expectation that most participants do not hold other people's vouchers indefinitely, but it is not treated as a universal behavioral law. Delay sensitivities change redemption timing. Probabilistic gating sensitivities change whether redemption occurs within a modeled opportunity. The sensitivity analysis compares voucher-to-voucher circulation, same-token return, stale voucher inventory, and route success under those timing and occurrence assumptions.

\subsection[Deployment monitoring priorities for productive-capacity assumptions]{Deployment monitoring priorities\\for productive-capacity assumptions}
The reported frontier depends on modeled productive-capacity feedback, stable-receipt timing, repayment behavior, and external stable-demand assumptions. The empirical data provide calibration evidence through pool value, verified reports, repayment proxies, stable receipts, and external stable-demand categories. The Monte Carlo uses these inputs as modeled productive-capacity assumptions rather than as causal income-growth estimates. Productive-capacity feedback means that a modeled credit event may later generate stable receipts or additional settlement capacity; it does not prove that every advance increases real income, welfare, ecological capacity, or fulfillment quality. Similarly, the simulator's consumer wallets and the empirical external non-producer stable-to-voucher category should not be collapsed into a single household-consumption interpretation. External non-producer demand is an observed stable-side purchase category, while consumer behavior in a deployment must be measured directly.

\begingroup
\small
\setlength{\tabcolsep}{3pt}
\begin{longtable}{P{0.24\textwidth}P{0.36\textwidth}P{0.30\textwidth}}
\caption{Deployment monitoring priorities for productive-capacity assumptions}
\label{tab:supp-deployment-monitoring-priorities}\\
\toprule
Monitoring domain & Why it matters for the frontier & Deployment measurement \\
\midrule
\endfirsthead
\toprule
Monitoring domain & Why it matters for the frontier & Deployment measurement \\
\midrule
\endhead
Producer productive return & The pass/fail grid is sensitive to whether borrowing produces later stable receipts rather than only more obligation stock. & Lag, size, volatility, and realized stable receipts after own-voucher borrowing or other accepted advances. \\
Stable retention versus exit & Stable receipts support repayment only if they can be used for obligation closure without unrealistic retention assumptions. & Share of producer stable receipts retained for repayment, spent outside the pool network, converted into other assets, or returned to pools. \\
Borrower self-repayment & Monthly stable-to-own-voucher repayment is a key channel for closing producer obligations and reducing pool exposure. & Stable-to-own-voucher repayment volume, arrears, deferred pressure, waiting balances, and time from advance to closure. \\
External non-producer demand & External stable demand helps remove obligations, but it is not automatically household consumption or guaranteed future demand. & Distinct buyers, repeat purchases, average ticket size, seasonality, volatility, and source classification. \\
Other-producer stable reuse & Producer stable-to-voucher reuse is empirically small and should remain separate from external-buyer demand. & Other-producer stable-to-voucher volume, frequency, counterparties, and share of producer stable receipts. \\
Voucher-to-voucher circulation & ROLA-like settlement quality can deteriorate even when scheduled bond service clears. & V2V value and share, active routable voucher float, same-token return, route success, and stale voucher inventory. \\
Borrowing-cap pressure & If producers become cap-bound, new liquidity may create queues or repayment waiting rather than productive circulation. & Cap utilization, cap-bound producer count, suppressed own-voucher borrowing, unresolved pool-held vouchers, and time to unlock limits. \\
Issuer service eligibility & Scheduled service should use only eligible recovered stable and cash-eligible fee resources, not generic turnover. & Recovered stable by source, fee conversion, lockbox eligibility, headroom after scheduled service, and excluded inventory turnover. \\
Household or participant cash stress & Repayment pressure must not become scarce-cash extraction from vulnerable participants. & Participant cash-stress indicators, complaint logs, repayment source surveys, coercion flags, and missed-essential-spending indicators. \\
Fulfillment and welfare evidence & Transaction closure is not evidence of welfare, ecological regeneration, or fulfillment quality by itself; verified reports provide component evidence for reported activities and should be linked to outcome protocols. & Verified reports, producer surveys, redemption satisfaction, repair completion, ecological evidence, social outcome indicators, and local governance review. \\
\bottomrule
\end{longtable}
\endgroup

These monitoring domains are falsification requirements for field deployment. If productive returns, external stable demand, redemption timing, or repayment closure are weaker than the calibrated assumptions used here, the settlement-capacity pass region should be expected to shrink. Conversely, stronger verified productive absorption, timely repayment, and non-coercive stable demand would widen the practical deployment region only after those effects are measured rather than assumed.

\subsection{Settlement velocity, routed volume, and fee-service capacity}
A useful design intuition behind a federated pool network is a velocity-of-settlement claim: reusable commitments can settle faster when they are accepted, routed, bought back, redeemed, netted, or repaired across more than one pool. This is related to the classical velocity equation, but the relevant stock is not only money. It is the stock of unsettled obligations. Settlement velocity concerns how quickly outstanding obligations close through redemption, buyback, routing, clearing, or repair. It is not ordinary transaction velocity and does not automatically create bond-service cash.

For voucher or obligation class $j$ measured over period $t$, define:
\begin{align*}
D_{jt} &= \text{average unsettled obligation stock},\\
S_{jt} &= S^{\mathrm{direct}}_{jt}+S^{\mathrm{routed}}_{jt}
       +S^{\mathrm{cleared}}_{jt}+S^{\mathrm{repair}}_{jt},\\
       &\quad \text{the obligation-closing settlement flow}.
\end{align*}
Then:
\[
\nu_{jt} = \frac{S_{jt}}{D_{jt}}
\]
is the class-specific settlement velocity. The ratio is undefined when $D_{jt}=0$ and should be reported as not applicable or handled by a pre-specified convention. It can be below, equal to, or above one depending on settlement flow relative to average unsettled stock. For a network of pools:
\[
\nu_t^{net} =
\frac{\sum_j S_{jt}}{\sum_j D_{jt}},
\qquad
\]
The network mechanism is that routing, clearing, and repair can increase the settlement-flow terms for a given stock of obligations. This is a claim about finding feasible settlement paths, not about relaxing issuer liability or ignoring pool limits.

Fee accounting requires a second distinction. Let $Q_t$ be gross routed or pool-executed value in period $t$, $\tau_t$ the effective fee rate on that value, and $\chi_t$ the cash-eligible or convertible share of fee inflows after policy and slippage constraints. Gross fee inflow is:
\[
\Phi_t^{gross} \approx \tau_t Q_t,
\]
and cash-eligible fee flow is:
\[
\Phi_t^{cash} \approx \chi_t \tau_t Q_t.
\]
If routed value is tightly linked to obligation closure, $Q_t$ may be approximated as $\lambda_t S_t=\lambda_t \nu_t D_t$, where $\lambda_t$ is the gross routed-value multiplier per unit of obligation-closing settlement. Then:
\[
\Phi_t^{cash} \approx \chi_t \tau_t \lambda_t \nu_t D_t.
\]
This approximation is useful for theory and sensitivity design, but it is not a result in the reported frontier. Gross route volume can rise because of useful settlement, because of multi-hop movement before settlement, or because of churn. Only realized cash-eligible fees and recovered stable assets that pass the bond-service rules enter scheduled service resources:
\[
E^B_t = H_t + \alpha_B \Phi_t^{cash} + R^{pay}_t.
\]
Thus higher settlement velocity can support infrastructure and debt-service capacity only when it corresponds to real obligation closure, acceptable routed value, cash-eligible fees, disclosed allocation rules, and reserves that are legally available for scheduled service. The reported Monte Carlo outputs already track swap volume, route success, fee-service reservation, recovered stable, scheduled-payment coverage, and headroom. A future reporting extension should add explicit $D_t$, $S_t$, $\nu_t$, $\lambda_t$, and $\chi_t$ diagnostics before making settlement velocity a headline empirical or frontier result.

\paragraph{Reproducibility status.}
Independent reproduction requires the archived simulator commit or release state, runner command, configuration file, random-seed policy, calibration input tables, privacy-safe derived input moments, and generated output files for each table and figure in the main paper and supplement. The publication package records LaTeX dependencies, aggregate calibration moments, reported summary outputs, checksums, package provenance, and ancillary notes under \texttt{anc/} where available. It is not a raw participant-level data archive; raw records are represented through the cited public anonymized dataset and privacy-safe derived outputs.

\subsection{Scenario map}
\begingroup
\footnotesize
\setlength{\tabcolsep}{2pt}
\begin{longtable}{P{0.05\textwidth}P{0.19\textwidth}P{0.17\textwidth}P{0.21\textwidth}P{0.26\textwidth}}
\caption{Monte Carlo scenario map and paper status}
\label{tab:supp-monte-carlo-scenarios}\\
\toprule
Sc. & Description & Purpose & Expected failure mode & Status in paper \\
\midrule
\endfirsthead
\toprule
Sc. & Description & Purpose & Expected failure mode & Status in paper \\
\midrule
\endhead
A & Conventional public bond spent through centralized procurement. & Baseline for local-settlement comparison. & Local commitments remain invisible. & Conceptual comparator only. \\
B & Green or social bond project finance with ordinary settlement architecture. & Separates use-of-proceeds improvement from settlement change. & Project value leaks externally. & Conceptual comparator only. \\
C & Mutual aid only, with ROLA-like support and no formal bond liquidity. & Baseline for local repair without capital injection. & Scale and liquidity constraints. & Empirical and conceptual baseline. \\
D & Sarafu-like local pools without formal bond layer. & Tests implemented pool curation, limits, swaps, and redemption. & Fragmentation and limited redemption paths. & Tested as full no-bond calibration fit and matched baseline. \\
E & Regenerative bond with gross principal deployed into eligible local commitment pools. & Core mechanism test. & Overvaluation, liquidity leakage, stable dependency, weak fulfillment, or settlement-attention capture. & Reported 64-cell settlement-capacity frontier. \\
F & Regenerative bond plus public procurement lane. & Tests demand anchor and local production retention. & Capture, political allocation, or bypass of producers. & Not reported as a result. \\
G & High coupon or high fixed operations cost. & Tests extraction risk, cash-headroom stress, and issuer sustainability. & Unpaid scheduled service or inadequate headroom. & Coupon stress tested; fixed-cost schedule not reported as a result. \\
H & Low convertibility: high in-kind fees but low cash-eligible share. & Tests cash coverage realism. & Operations or scheduled-service shortfall despite volume. & Diagnostic only. \\
I & Governance capture: central actor overrides local limits or listings. & Tests anti-capture design. & Volume rises while trust, fulfillment, and repair fall. & Not reported as a result. \\
\bottomrule
\end{longtable}
\endgroup

\begin{table}[htbp]
\centering
\caption{Reported settlement-capacity frontier by coupon}
\label{tab:supp-reported-frontier-by-coupon}
\small
\setlength{\tabcolsep}{4pt}
\begin{tabularx}{\textwidth}{YYYYY}
\toprule
Annual coupon & Max pass principal ratio & Max strong principal ratio & First failed principal ratio & Binding failure mode \\
\midrule
0\% & 2.00 & 2.00 & -- & None in grid \\
4\% & 2.00 & 1.50 & -- & None in grid \\
8\% & 2.00 & 1.50 & -- & None in grid \\
12\% & 1.50 & 1.50 & 2.00 & Scheduled payment coverage \\
15\% & 1.50 & 1.00 & 2.00 & Scheduled payment coverage \\
25\% & 1.50 & 1.00 & 2.00 & Scheduled payment coverage \\
35\% & 1.00 & 1.00 & 1.50 & Scheduled payment coverage \\
45\% & 1.00 & 0.75 & 1.50 & Scheduled payment coverage \\
\bottomrule
\end{tabularx}
\end{table}

Principal ratio is gross bond principal divided by modeled eligible backing capacity under the reported eligibility policy. The frontier uses a 260-week term, 26-week issuer-payment stride, even principal amortization across payment periods, coupon accrual on start-of-period outstanding principal, zero issuer reserve share, and a remaining-schedule bond-service lockbox. These are simulator schedule assumptions and stress-test settings, not proposed securities terms. The reported 100-run-per-cell design supports the paper's frontier classification for the specified runner, configuration, and seed policy; p05 and p50 values are simulation guardrail summaries rather than estimates of universal legal, financial, or deployment probabilities.

The table reports the current frontier under the stated topology, fee-service share, and operating-cost exclusions. It is not a general bond-sizing rule. Lower cash-eligible fee share, delayed third-party redemption, weaker route visibility, explicit fixed operating costs, insurance claims, legal/tax costs, or governance override stress would be expected to shrink the maximum pass and strong principal ratios. These sensitivities are design priorities for the next validation stage; the present paper finalizes the reported frontier as a calibrated stress-test result, not as an optimized financing recommendation.

\subsection{Additional frontier figure}

\begin{figure}[htbp]
\centering
\includegraphics[width=0.9\textwidth]{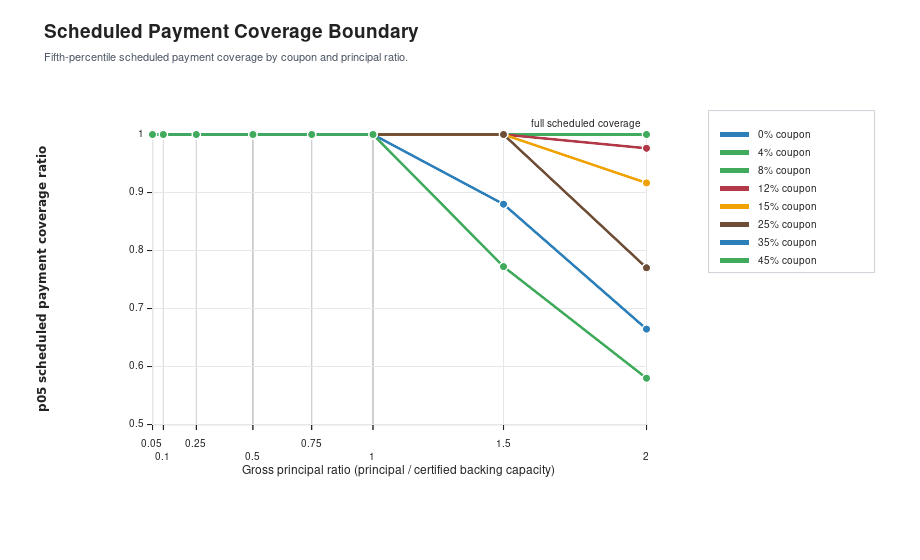}
\caption{Final settlement-capacity grid: scheduled-payment coverage boundary. Higher principal and coupon stress lower fifth-percentile scheduled-payment coverage, making formal service coverage the clearest first boundary in the tested grid.}
\label{fig:supp-edge-diagnostic-payment-boundary}
\end{figure}

The main article reports the outcome grid and the simulated settlement motif-share diagnostic. Figure~\ref{fig:supp-edge-diagnostic-payment-boundary} reports the complementary scheduled-payment coverage boundary.

\FloatBarrier

\section{Calibration Tables}
\begin{table}[htbp]
\centering
\caption{Reported paper-cohort activity and report moments used by the calibrated runner}
\label{tab:supp-sarafu-calibrated-tier-summary}
\small
\setlength{\tabcolsep}{3pt}
\begin{tabularx}{\textwidth}{P{0.13\textwidth}YYYY}
\toprule
Tier & Pools & Swap events & Verified report exposure & Backing inflow \\
\midrule
Strong & 45 & 5,535 & 1,124 & 17,249.59 \\
Moderate & 21 & 2,385 & 912 & 5,663.15 \\
Weak & 7 & 496 & 627 & 3,933.48 \\
\bottomrule
\end{tabularx}
\end{table}

\begin{table}[htbp]
\centering
\caption{Reported return, borrowing-closure, and backing moments used by the engine calibration-fit gate}
\label{tab:supp-sarafu-calibrated-tier-settlement}
\small
\setlength{\tabcolsep}{3pt}
\begin{tabularx}{\textwidth}{P{0.13\textwidth}YYYYY}
\toprule
Tier & Same-token return & Cash return & Voucher return & Borrow return & Backing inflow \\
\midrule
Strong & 0.391 & 0.199 & 0.631 & 0.399 & 17,249.59 \\
Moderate & 0.537 & 0.315 & 0.928 & 0.646 & 5,663.15 \\
Weak & 0.189 & 0.120 & 0.859 & 0.424 & 3,933.48 \\
\bottomrule
\end{tabularx}
\end{table}

\begin{table}[htbp]
\centering
\caption{Reported cohort activity and report targets}
\label{tab:supp-sarafu-baseline-validation}
\small
\setlength{\tabcolsep}{3pt}
\begin{tabularx}{\textwidth}{P{0.13\textwidth}YYP{0.10\textwidth}}
\toprule
Tier & Pools & Swap events & Verified report exposure \\
\midrule
Strong & 45 & 5,535 & 1,124 \\
Moderate & 21 & 2,385 & 912 \\
Weak & 7 & 496 & 627 \\
\bottomrule
\end{tabularx}
\end{table}

\begin{table}[htbp]
\centering
\caption{Reported cohort settlement and backing targets}
\label{tab:supp-sarafu-baseline-settlement-validation}
\small
\setlength{\tabcolsep}{3pt}
\begin{tabularx}{\textwidth}{P{0.13\textwidth}YYP{0.10\textwidth}}
\toprule
Tier & Same-token return & Backing inflow & Borrow return \\
\midrule
Strong & 0.391 & 17,249.59 & 0.399 \\
Moderate & 0.537 & 5,663.15 & 0.646 \\
Weak & 0.189 & 3,933.48 & 0.424 \\
\bottomrule
\end{tabularx}
\end{table}
\FloatBarrier

\section{Background Template Counterfactuals}
The following Sarafu-calibrated template counterfactuals are retained as background evidence about repayable-liquidity cashflow and allocation tradeoffs. They are not the reported issuer and eligible-pool repayment model.

\begin{table}[htbp]
\centering
\caption{Sarafu-calibrated counterfactual return amounts at central coupon target}
\label{tab:supp-sarafu-counterfactual-results}
\small
\setlength{\tabcolsep}{3pt}
\begin{tabularx}{\textwidth}{P{0.35\textwidth}YYY}
\toprule
Policy & Principal & Coupon due & Fee return \\
\midrule
Aid/grant baseline & 0 & 0 & 0 \\
Bond broad equal & 400,000 & 24,000 & 131,416 \\
Bond strong-pool activity & 400,000 & 24,000 & 134,163 \\
Bond weak-capacity inclusion & 400,000 & 24,000 & 131,500 \\
\bottomrule
\end{tabularx}
\end{table}

\begin{table}[htbp]
\centering
\caption{Sarafu-calibrated counterfactual return ratios at central coupon target}
\label{tab:supp-sarafu-counterfactual-return-ratios}
\small
\setlength{\tabcolsep}{3pt}
\begin{tabularx}{\textwidth}{P{0.45\textwidth}YY}
\toprule
Policy & Coupon coverage & Principal recovery \\
\midrule
Aid/grant baseline & N/A & N/A \\
Bond broad equal & 5.48 & 0.329 \\
Bond strong-pool activity & 5.59 & 0.335 \\
Bond weak-capacity inclusion & 5.48 & 0.329 \\
\bottomrule
\end{tabularx}
\end{table}

The aid/grant baseline has zero principal and zero coupon due, so coupon coverage and principal recovery are undefined rather than equal to zero or one.

\begin{table}[htbp]
\centering
\caption{Sarafu-calibrated counterfactual stress diagnostics at central coupon target}
\label{tab:supp-sarafu-counterfactual-stress-diagnostics}
\small
\setlength{\tabcolsep}{3pt}
\begin{tabularx}{\textwidth}{P{0.45\textwidth}YY}
\toprule
Policy & Leakage & Repayment closure \\
\midrule
Aid/grant baseline & 0.000 & 0.470 \\
Bond broad equal & 0.861 & 0.451 \\
Bond strong-pool activity & 0.769 & 0.556 \\
Bond weak-capacity inclusion & 0.862 & 0.458 \\
\bottomrule
\end{tabularx}
\end{table}
\FloatBarrier

\section{Glossary}
\begingroup
\small
\setlength{\tabcolsep}{4pt}
\begin{longtable}{P{0.24\textwidth}P{0.68\textwidth}}
\caption{Selected terms}
\label{tab:supp-glossary}\\
\toprule
Term & Plain-language meaning \\
\midrule
\endfirsthead
\toprule
Term & Plain-language meaning \\
\midrule
\endhead
Coupon & Interest or other periodic return due from a bond issuer to investors under bond terms. \\
Principal & The amount borrowed through the bond and due at maturity or according to the amortization schedule. \\
Regenerative bond & A formal debt instrument whose disclosed use-of-proceeds and governance rules allocate proceeds to locally governed settlement systems designed to strengthen settlement capacity across locally specified productive, ecological, care, mutual-aid, and repair commitments. Scheduled obligations to bondholders remain with the formal issuer; local pools, households, stewards, and redeemable commitments are not bondholder collateral unless explicitly and legally structured otherwise. \\
Commitment pool & A locally governed arrangement that curates, values, limits, and exchanges redeemable commitments, stable-credit claims, or vouchers. \\
Community Asset Voucher & A redeemable commitment issued by a person, group, or institution for goods, services, labor, or other accepted fulfillment. \\
Stable or stable credit & A cash-like or stable-value settlement asset used in the data or simulator; it is not treated here as legal tender, a bank deposit, an accounting cash equivalent, or a central-bank liability. \\
ROLA-like & A behavioral proxy for rotating-labor association motifs represented in commitment pools as movable reciprocal commitments, especially goods, services, labor, care, ecological work, voucher-to-voucher exchange, routing, commitment-issuer return, redemption, clearing, or repair. \\
ROSCA-like & A behavioral proxy for rotating savings and credit association motifs represented in commitment pools as stable or cash-like deposits, borrowing rights, credit limits, repayment obligations, mature borrow-proxy closure, and stable recovery. \\
Route graph & A network of pools where settlement paths may move across more than one pool under local rules. \\
Same-token return & A diagnostic asking whether an asset that leaves a pool later returns to that same pool. \\
Voucher-to-voucher value & Stable-value-denominated value of voucher-for-voucher exchange through pools; in the reported frontier specification it is the main ROLA-like value-preservation and amplification metric. \\
Material-decline threshold & Pre-specified tolerance for how much a protected metric, such as V2V value, may fall relative to its matched no-bond baseline before the cell is treated as degraded. The reported V2V value and active-float thresholds are 15 percent declines relative to matched no-bond baselines. \\
Active-float guardrail & Pre-specified test that the modeled stock of active circulating or settlement-relevant vouchers remains within an acceptable range relative to baseline. The reported active-float guardrail tests active routable producer-voucher float against a 15 percent decline threshold. \\
p05 / p50 & Monte Carlo summary statistics: p05 is the fifth percentile across runs and p50 is the median across runs. \\
Debt-removal proxy & A stable-to-voucher pattern followed by commitment-issuer return or burn, interpreted as evidence that stable liquidity reduced or closed a voucher obligation. \\
Bond-service lockbox & Segregated accounting account that holds eligible recovered stable or cash-like value for scheduled issuer debt service; it is not unrestricted issuer headroom unless release conditions are met. \\
Issuer operating/risk headroom & Modeled resources above the scheduled amount due to bondholders; in the reported frontier this is summarized as a median headroom multiple, with 1.25 meaning at least 25 percent above scheduled service before full costs. It is candidate capacity, not guaranteed profit or an additional bondholder claim. \\
Productive-capacity feedback & A modeled pathway in which successful modeled producer credit is assumed to increase later productive inflow, subject to calibration caps and matched no-bond comparison. \\
\bottomrule
\end{longtable}
\endgroup

\section{Technical Source Integration}
Project technical materials inform implementation design, but they are not independent evidence that regenerative bonds perform as intended.

\begingroup
\small
\setlength{\tabcolsep}{3pt}
\begin{longtable}{P{0.22\textwidth}P{0.29\textwidth}P{0.22\textwidth}P{0.18\textwidth}}
\caption{Technical sources and manuscript use}
\label{tab:supp-technical-source-integration}\\
\toprule
Source & Relevant content & Paper use & Caution \\
\midrule
\endfirsthead
\toprule
Source & Relevant content & Paper use & Caution \\
\midrule
\endhead
Cosmo-Local Credit documentation \citep{cosmoLocalCreditDocs}
& Clearing network for redeemable commitments, curated markets, governance, and resilient settlement.
& Mechanism vocabulary and protocol mapping.
& Design vocabulary only; empirical and legal claims require independent evidence. \\
\addlinespace
CLC protocol repository \citep{cosmoLocalCreditProtocol}
& Community asset vouchers, commitment pools, fees, route logic, registry, limiters, fee policy, quoters, splitter, and router.
& Software invariant and implementation feasibility.
& Implements mechanics without establishing legal compliance or economic performance. \\
\addlinespace
CLC simulator repository \citep{cosmoLocalCreditSim}
& Agent-owned pools, vouchers, stable assets, liquidity providers, producers, consumers, routing, clearing, fees, waterfall, insurance incidents, and cash-eligible assets.
& Computational substrate for Monte Carlo regenerative-bond simulation.
& The issuer, service, governance, ecological, and Sarafu-calibration layers are specified in the model reported here. \\
\addlinespace
Sarafu Network repository and dataset \citep{sarafuNetworkRepo,ruddick2026sarafuDataset}
& Public anonymized transaction, pool, voucher, limit, report, and activity data from an implemented system.
& Empirical calibration and component evidence for commitment-pool mechanics.
& Public anonymization still requires data ethics, aggregation choices, and careful inference. \\
\bottomrule
\end{longtable}
\endgroup

\section{Core Simulation Variables}
\begingroup
\small
\setlength{\tabcolsep}{3pt}
\begin{longtable}{P{0.14\textwidth}P{0.36\textwidth}P{0.20\textwidth}P{0.20\textwidth}}
\caption{Core simulation variables and calibration paths}
\label{tab:supp-simulation-variables}\\
\toprule
Variable & Meaning & Unit & Calibration path \\
\midrule
\endfirsthead
\toprule
Variable & Meaning & Unit & Calibration path \\
\midrule
\endhead
$B_0$ & Bond principal raised. & Currency unit. & Scenario parameter. \\
$c$ & Coupon rate. & Annual percent. & Scenario parameter. \\
$DS_t$ & Principal and coupon due in period $t$. & Currency unit. & Bond terms. \\
$R^{pay}_t$ & Reserve amount legally available to pay scheduled service in period $t$. & Currency unit. & Scenario state and instrument terms. \\
$R^{rel}_t$ & Reserve amount released for unrestricted issuer use after scheduled service and required reserve targets are met. & Currency unit. & Scenario state and release rules. \\
$LP_t$ & Local-pool liquidity allocation. & Currency unit. & Sarafu injection motifs and scenario policy. \\
$H_t$ & Eligible recovered pool stable available for scheduled service. & Currency unit. & Producer repayment, stable purchases, maturity settlement, and clearing diagnostics. \\
$X^B_t$ & Gross service headroom above scheduled due before issuer operating and risk costs. & Currency unit. & Frontier cash-headroom diagnostic. \\
$\eta^B_t$ & Headroom coverage multiple used for the reported p50 guardrail. & Ratio. & Modeled service-and-release resources divided by scheduled service. \\
$D_{at}$ & Outstanding commitments of commitment issuer $a$. & Common value index. & Voucher outstanding and pool inventories. \\
$S_{at}$ & Settlements or commitment-issuer-return proxies. & Common value index per period. & Redemptions, burns, commitment-issuer returns. \\
$\chi_t$ & Cash-eligible fee share. & Ratio. & Fee asset mix and conversion rules. \\
$\alpha_B$ & Published allocation share of cash-eligible fee flow or other eligible flow assigned to scheduled bond service. & Ratio. & Scenario policy and lockbox rule. \\
$\lambda_t$ & Gross routed-value multiplier per unit of obligation-closing settlement. & Ratio. & Derived diagnostic for settlement-velocity sensitivity; not a headline reported frontier result. \\
$D_t$ & Average unsettled obligation stock used for settlement-velocity diagnostics. & Common value index. & Voucher inventory, active producer debt, and unsettled pool exposure. \\
$S_t$ & Obligation-closing settlement flow. & Common value index per period. & Redemptions, buybacks, same-token returns, circulation closure, clearing, and repair. \\
$\nu_t$ & Settlement velocity, $S_t/D_t$. & Flow/stock ratio. & Derived diagnostic; not a headline reported frontier result. \\
$Q_t$ & Gross routed or pool-executed value. & Common value index per period. & Swap and route volume; may exceed obligation-closing settlement. \\
\bottomrule
\end{longtable}
\endgroup

\bibliographystyle{achicago}
\bibliography{regenerative_bonds}